\documentclass[12pt,letter]{article}
\pdfoutput=1
\usepackage{graphicx, epsfig, color,cite}
\usepackage{amsmath}
\usepackage{amssymb}
\usepackage{caption,subcaption,graphicx}
\usepackage[hidelinks]{hyperref}
\def\to{\rightarrow}

\def\bi{\begin{itemize}}
\def\ei{\end{itemize}}

\def\tchi{\tilde\chi}
\def\tm{\tilde m}

\def\tst{\tilde t}

\def\tg{\tilde g}

\def\tw{\widetilde\chi^{\pm}}
\def\tz{\widetilde\chi^0}
\def\alt{\lesssim}
\def\agt{\gtrsim}
\def\be{\begin{equation}}  
\def\ee{\end{equation}}  
\def\bea{\begin{eqnarray}}  
\def\eea{\end{eqnarray}}

\begin{document}
\begin{titlepage}
\begin{flushright}
OU-HEP-190617
\end{flushright}

\vspace{0.5cm}
\begin{center}
{\Large \bf Naturalness versus stringy naturalness\\
(with implications for collider and dark matter searches)
}\\ 
\vspace{1.2cm} \renewcommand{\thefootnote}{\fnsymbol{footnote}}
{\large Howard Baer$^1$\footnote[1]{Email: baer@ou.edu },
Vernon Barger$^2$\footnote[2]{Email: barger@pheno.wisc.edu} and
Shadman Salam$^1$\footnote[3]{Email: shadman.salam@ou.edu}
}\\ 
\vspace{1.2cm} \renewcommand{\thefootnote}{\arabic{footnote}}
{\it 
$^1$Homer L. Dodge Department of Physics and Astronomy,
University of Oklahoma, Norman, OK 73019, USA \\[3pt]
}
{\it 
$^2$Department of Physics,
University of Wisconsin, Madison, WI 53706 USA \\[3pt]
}

\end{center}

\vspace{0.5cm}
\begin{abstract}
\noindent
The notion of stringy naturalness-- that an observable ${\cal O}_2$ is more natural than 
${\cal O}_1$ if more (phenomenologically acceptable) vacua solutions lead to ${\cal O}_2$ 
rather than ${\cal O}_1$-- is examined within the context of the Standard Model (SM) and
various SUSY extensions: CMSSM/mSUGRA, high-scale SUSY and radiatively-driven natural SUSY (RNS).
Rather general arguments from string theory suggest a (possibly mild) statistical draw towards 
vacua with large soft SUSY breaking terms. 
These vacua must be tempered by an anthropic veto of non-standard vacua or 
vacua with too large a value of the weak scale $m_{weak}$.
We argue that the SM, the CMSSM and various high-scale SUSY models are all 
expected to be relatively rare occurances within the string theory landscape of vacua.
In contrast, models with TeV-scale soft terms but with $m_{weak}\sim 100$ GeV 
and consequent light higgsinos (SUSY with radiatively-driven naturalness)
should be much more common on the landscape.
These latter models have a statistical preference for $m_h\simeq 125$ GeV and 
strongly interacting sparticles beyond current LHC reach. 
Thus, while conventional naturalness favors sparticles close to the weak scale, 
stringy naturalness favors sparticles so heavy that electroweak symmetry is barely broken and one
is living dangerously close to vacua with charge-or-color breaking minima, no electroweak
breaking or pocket universe weak scale values too far from our measured value.
Expectations for how landscape SUSY would manifest itself at collider and dark matter search 
experiments are then modified compared to usual notions.
\end{abstract}
\end{titlepage}

\section{Introduction}
\label{sec:intro}

{\it Naturalness:} The gauge hierarchy problem (GHP)\cite{ghp}-- 
what stabilizes the weak scale so that it doesn't blow up 
to the GUT/Planck scale-- 
is one of the central conundrums of particle physics. 
Indeed, it provides crucial motivation for the premise that new physics 
should be lurking in or around the weak scale. 
Supersymmetric models (SUSY) with weak scale soft SUSY breaking terms 
provide an elegant solution to the GHP\cite{witten,WSS} 
but so far weak-scale sparticles have failed to appear at the 
CERN Large Hadron Collider (LHC) and WIMPs have failed to appear in
direct detection experiments\cite{Xe1ton}.
The latest search limits from LHC Run 2 require gluinos with 
$m_{\tg}\agt 2.25$ TeV\cite{lhc_gl} 
and top-squarks $m_{\tst_1}\agt 1.1$ TeV\cite{lhc_t1}. 
These lower bound search limits stand in sharp contrast to early 
sparticle mass upper bounds from naturalness that seemingly required 
$m_{\tg,\tst_1}\alt 0.4$ TeV\cite{EENZ,BG,DG,AC}.\footnote{
For a recent analysis in the context of EENZ/BG, see {\it e.g.} 
Ref. \cite{Kim:2013uxa}.}
Thus, LHC limits seem to imply the soft SUSY breaking scale $m_{soft}$ 
lies in the
multi-TeV rather than the weak-scale range. 
This then opens up a Little Hierarchy Problem (LHP)\cite{BS}:
why does the weak scale not blow up to the energy scale associated 
with soft SUSY breaking, {\it i.e.} why is $m_{weak}\ll m_{soft}$?

Early upper bounds on sparticle masses derived from naturalness 
were usually computed using the EENZ/BG log-derivative measure\cite{EENZ,BG}: 
for an observable ${\cal O}$, then 
\begin{equation}
\Delta_{BG}({\cal O})\equiv {\rm max}_i|\frac{\partial\log{\cal O}}{\partial\log p_i}|= {\rm max}_i|\frac{p_i}{\cal O}\frac{\partial{\cal O}}{\partial p_i}|
\label{eq:DBG}
\end{equation}
where the $p_i$ are fundamental parameters of the underlying theory. 
For an observable depending linearly on model parameters, 
${\cal O}=a_1p_1+\cdots +a_np_n$, 
then $\partial{\cal O}/\partial p_i=a_i$ and $\Delta_{BG}({\cal O})$ just picks off 
the maximal right-hand-side contribution to ${\cal O}$ and compares it to ${\cal O}$.
In the case where one contribution $a_ip_i\gg {\cal O}$, then some other 
contribution(s) will have to be finetuned to large opposite-sign values 
such as to maintain ${\cal O}$ at its measured value. 
Such finetuning of fundamental parameters seems highly implausible in nature
absent some symmetry or parameter selection mechanism.
Thus, the log-derivative is a measure of\cite{land3}: 
\begin{quotation}
{\it Practical naturalness}:
An observable ${\cal O}$ is natural if all {\it independent} contributions to 
${\cal O}$ are comparable to or less than ${\cal O}$.
\end{quotation}

For the case of the LHP, the observable ${\cal O}$ is traditionally take to be 
$m_Z^2$ and the $p_i$ are taken to be the MSSM $\mu$ term and 
soft SUSY breaking terms so that naturalness then requires all 
contributions to $m_Z^2$ to be comparable to $m_Z^2$: 
this is the basis for expecting sparticles to occur around the 100 GeV scale.
A fundamental issue in computing $\Delta_{BG}$ is\cite{dew,mt,seige,arno}: 
what is the correct choice to be made for the $p_i$? 
If soft terms are correlated with one another, as expected in fundamental 
supergravity/superstring theories, then one gets a very different answer 
than if one assumes some effective $4-d$ SUSY theory with many 
independent soft parameters which are introduced to parametrize one's 
ignorance of their origin.\footnote{For instance, 
in dilaton-dominated SUSY breaking, one expects $m_0^2=m_{3/2}^2$ and 
$m_{1/2}=-A_0=\sqrt{3}m_{3/2}$. In such a case, 
it would not make sense to adopt $m_0$ and $m_{1/2}$ as  free, independent 
parameters.} 
Alternatively, even if the parameters $p_i$ are independent, it is possible 
that some selection mechanism is responsible for the values towards which they tend.

It is also common in the literature to apply practical naturalness to the 
Higgs mass: 
\begin{equation}
m_h^2\simeq m_{H_u}^2(weak)+\mu^2(weak)+mixing+rad.\ corr.
\label{eq:mhs}
\end{equation}
where the mixing and radiative corrections are both comparable to $m_h^2$. 
Also, $m_{H_u}^2(weak)=m_{H_u}^2(\Lambda )+\delta m_{H_u}^2$
where it is common to estimate $\delta m_{H_u}^2$ using its renormalization
group equation (RGE) by setting several terms in $dm_{H_u}^2/dt$
(with $t=\log Q^2$) to zero so as to integrate in a single step: 
\begin{equation}
\delta m_{H_u}^2\sim -\frac{3f_t^2}{8\pi^2}(m_{Q_3}^2+m_{U_3}^2+A_t^2)\ln\left( \Lambda^2/m_{soft}^2\right).
\end{equation}
Taking $\Lambda\sim m_{GUT}$ and requiring the high scale measure 
\begin{equation}
\Delta_{HS}\equiv\delta m_{H_u}^2/m_h^2
\label{eq:DHS}
\end{equation}
$\Delta_{HS}\alt 1$ then requires three third generation squarks lighter 
than 500 GeV\cite{prw,bkls} (now highly excluded by LHC top-squark searches) 
and small $A_t$ terms (whereas $m_h\simeq 125$ GeV typically requires 
large mixing and thus multi-TeV values of $A_0$\cite{mhiggs,h125}). 
The simplifications made in this calculation ignore the fact that
$\delta m_{H_u}^2$ is highly dependent on $m_{H_u}^2(\Lambda )$ 
(which is set to zero in the simplification)\cite{dew,seige,arno}. 
In fact, the larger one makes $m_{H_u}^2(\Lambda )$, 
then the larger becomes the cancelling correction $\delta m_{H_u}^2$. 
Thus, these terms are {\it not independent}: one cannot tune $m_{H_u}^2(\Lambda)$ 
against a large contribution $\delta m_{H_u}^2$. 
Thus, weak-scale top squarks and small $A_t$ are not required by naturalness.

To ameliorate the above naturalness calculational quandaries, 
a more model independent measure $\Delta_{EW}$ was introduced\cite{ltr,rns}.\footnote{
A desirable feature of $\Delta_{EW}$ is that for a given SUSY spectrum, one obtains
exactly the same finetuning measure whether the spectrum is generated from multi- or
few parameter theory or at the weak scale (such as pMSSM) or at much higher scales.
This model independence is not shared by other measures such as $\Delta_{BG}$ or $\Delta_{HS}$.}
By minimizing the weak-scale SUSY Higgs potential, including radiative 
corrections, one may relate the measured value of the $Z$-boson mass 
to the various SUSY contributions:
\bea
m_Z^2/2 &=&\frac{m_{H_d}^2+\Sigma_d^d-(m_{H_u}^2+\Sigma_u^u)\tan^2\beta}{\tan^2\beta -1}-\mu^2\\ \nonumber
&\simeq &-m_{H_u}^2-\mu^2-\Sigma_u^u(\tst_{1,2}) .
\label{eq:mzs}
\eea
The measure 
\begin{equation}
\Delta_{EW}=|(max\ RHS\ contribution)|/(m_Z^2/2)
\label{eq:DEW}
\end{equation} 
is then low provided all {\it weak-scale} contributions to $m_Z^2/2$ 
are comparable to or less than $m_Z^2/2$. 
The $\Sigma_u^u$ and $\Sigma_d^d$ contain over 40 radiative corrections 
which are listed in the Appendix of Ref.~\cite{rns}. 
The conditions for natural SUSY (for {\it e.g.} $\Delta_{EW}<30$)\footnote{
The onset of finetuning for $\Delta_{EW}\agt 30$ is visually displayed in 
Fig. 1 of Ref. \cite{upper}.} 
can then be read off from Eq. \ref{eq:mzs}:
\begin{itemize}
\item The superpotential $\mu$ parameter has magnitude not too far from the 
weak scale, $|\mu |\alt 300$ GeV. 
This implies the existence of light higgsinos $\tz_{1,2}$ and
$\tw_1$ with $m(\tz_{1,2},\tw_1)\sim 100-300$ GeV.
\item $m_{H_u}^2$ is radiatively driven from large high scale values to 
small negative values at the weak scale 
(SUSY with radiatively-driven naturalness or RNS\cite{ltr}).
\item Large cancellations occur in the $\Sigma_u^u (\tst_{1,2})$ terms for large 
$A_t$ parameters which then allow for $m_{\tst_1}\sim 1-3$ TeV 
for $\Delta_{EW}<30$. 
The large $A_t$ term also lifts the Higgs mass $m_h$ into the vicinity of 125 GeV.
The gluino contributes to the weak scale at two-loop order so its mass can 
range up to $m_{\tg}\alt 6$ TeV with little cost to naturalness\cite{rns,upper,jamie}.
\item Since first/second generation squarks and sleptons contribute to the 
weak scale through (mainly cancelling) $D$-terms, they can range up to 
10-30 TeV at little cost to naturalness 
(thus helping to alleviate the SUSY flavor and CP problems)\cite{maren}.
\end{itemize}
By combining dependent soft terms in $\Delta_{BG}$ or by combining the dependent 
terms $m_{H_u}^2(\Lambda )$ and $\delta m_{H_u}^2$ in $\Delta_{HS}$, 
then these measures roughly reduce to $\Delta_{EW}$ 
(aside from the radiative contributions $\Sigma_{u,d}^{u,d}$).
Since $\Delta_{EW}$ is determined by the weak scale SUSY parameters, 
then different models which give rise to exactly the same sparticle mass
spectrum will have the same finetuning value (model independence).
Using the naturalness measure $\Delta_{EW}$, then it has been shown
that plenty of SUSY parameter space remains natural even in the face of
LHC Run 2 Higgs mass measurements and sparticle mass limits\cite{rns}.

{\it String theory landscape}: 
Weinberg's anthropic solution to the cosmological constant (CC)\cite{Weinberg:1987dv}, 
along with the emergence of the string theory landscape of vacua\cite{BP,Susskind:2003kw}, 
have presented a challenge to the usual notion of naturalness. 
In Weinberg's view, in the presence of a vast assortment ($\gg 10^{120}$) 
of pocket universes which are part of an eternally inflating multiverse, 
it may not be so surprising to find ourselves in one with 
$\Lambda_{cc}\sim 10^{-120}m_P^4$ since if it were
much larger, then the expansion rate would be so large that galaxies would not
be able to condense, and life as we know it would be unable to emerge.
This picture was bolstered by the discovery of the string theory
discretuum of flux vacua\cite{BP,Susskind:2003kw} where each metastable minimum of the 
string theory scalar potential would have a different value of $\Lambda_{cc}$,
and also different laws of physics and perhaps even different spacetime 
dimensions. The number of metastable minima has been estimated at around
$10^{500}$\cite{Ashok:2003gk} although far larger numbers have also been considered.
In such a scenario, then it is possible that many of the constants of nature
may take on environmentally determined values rather than being determined by
fundamental underlying principles.

One example of the challenge to naturalness comes in the form of 
Split Supersymmetry (SS)\cite{ArkaniHamed:2004fb}. 
In SS, one retains the positive features of 
gauge coupling unification and a WIMP dark matter candidate while eschewing the
motivation of naturalness. Then the expected SUSY particle spectrum of SS
can contain weak scale gauginos and higgsinos (which furnish gauge coupling 
unification and a WIMP dark matter candidate) while allowing most scalars
(except the SM-like Higgs doublet) to gain unnatural mass values
of perhaps $\tm\sim 10^3-10^{8}$ TeV. Here, one might expect the weak scale 
to blow up to the $\tm$ scale, but the anthropic requirement of $m_{weak}\sim 100$ GeV
selects a finely-tuned scalar spectrum that would otherwise seem unnatural.
In a similar vein, the notion of high-scale SUSY has also been entertained
wherein all sparticle masses are large and unnatural, perhaps at the
$10^2-10^3$ TeV scale\cite{HSSUSY,Wells:2004di,Hall:2011jd,Arvanitaki:2012ps}.

At first glance, one might expect that the string landscape would make BSM
physics non-predictive since: 
how are we to determine the metastable vacuum minimum that our universe 
inhabits out of $\sim 10^{500}$ or more choices?
To make progress, Douglas and collaborators\cite{Douglas:2004zg} have advanced the notion of 
a {\it statistical} program for determining BSM physics. 
In this case, if we can identify statistical trends for the many landscape 
vacua solutions, then we might be able to determine probabilistically 
what sort of pocket universe we are likely to live in.

To this end, Douglas has proposed the notion of {\it stringy naturalness}\cite{Douglas:2004zg}:
\begin{quotation}
{\bf Stringy naturalness:} the value of an observable ${\cal O}_2$ 
is more natural than a value ${\cal O}_1$ if more 
{\it phenomenologically viable} vacua lead to  ${\cal O}_2$ than to ${\cal O}_1$.
\end{quotation}
If we apply this definition to the cosmological constant, then 
{\it phenomenologically viable} is interpreted in an anthropic context in that we
must veto vacua which would not allow galaxy condensation. Out of the remaining 
viable vacua, we would expect $\Lambda_{cc}$ to be nearly as large as 
anthropically possible since there is more volume in $\Lambda_{cc}$ space
for larger values. Such reasoning allowed Weinberg to predict the value of
$\Lambda_{cc}$ to within a factor of a few of its measured value more than 
a decade before its value was determined from experiment\cite{Weinberg:1987dv}.

Our goal in this paper is to examine Douglas' notion of stringy naturalness
and to compare and contrast it to the above conventional notions of naturalness. 
We will examine what naturalness and what stringy naturalness imply for
the SM, the scale of SUSY breaking in SUSY models, and for the magnitude of the weak scale.
Our central conclusion from stringy naturalness is: 
{\it the soft SUSY breaking terms should be as large as
possible subject to the constraint that the value of the weak scale 
in various pocket universes-- with the MSSM as the low energy effective theory--
not deviate by more than a factor of a few from its measured value}.
This is in contrast to the conventional measures which tend to favor 
smaller soft terms comparable to the weak scale. 
In fact, stringy naturalness in this form with a relatively mild draw 
to large soft terms statistically favors a light 
Higgs mass $m_h\sim 125$ GeV with as yet no sign of sparticles at LHC.

In Sec. \ref{sec:lsusy}, we present details of how to implement the notion of stringy 
naturalness including Douglas' notion of a prior distribution of power-law
probability increase in soft term values. This is to be combined with a
{\it selection} criteria enforcing that the value of the weak scale in 
various pocket universes not deviate from our measured value by a factor 
of a few. The latter notion has been presented in some detail by 
calculations of Agrawal {\it et al.}\cite{Agrawal:1997gf}.
The landscape selection of the cosmological constant, 
as emphasized by Douglas {\it et al.}\cite{Douglas:2004qg}, 
operates separately from the soft SUSY breaking term/weak scale 
selection criteria.

In Sec. \ref{sec:sm}, we examine first what stringy naturalness implies for
the (non-SUSY) Standard Model (SM). 
We find  the SM-- valid up to a cutoff scale $\Lambda_{SM}$-- 
to be highly improbable within the landscape for a cutoff 
$\Lambda_{SM}\gg m_{weak}$. 
In Sec. \ref{sec:cmssm}, we argue that the paradigm CMSSM SUSY model should 
also be relatively rare on the landscape.
In Sec. \ref{sec:rns}, we examine the case of the MSSM on the landscape. 
In this case, there can be significant 
swaths of model parameters leading to SUSY with radiatively driven 
naturalness (RNS).
In fact, the statistical draw to large soft terms coupled with a weak scale
not too far from its measured value, is exactly what is needed for SUSY
with radiatively-driven naturalness.
In Sec. \ref{sec:hsSUSY}, we present arguments as to why
unnatural SUSY models such as split SUSY, high scale SUSY, 
spread SUSY and minisplit SUSY are likely to be relatively rare occurances on the
landscape. 
In Sec. \ref{sec:collDM}, we briefy discuss consequences of stringy naturalness for
collider and dark matter searches. A summary and our conclusions are presented
in Sec. \ref{sec:conclude}.

The present work is a continuation of a theme started in 
Ref.~\cite{Baer:2016lpj} where a qualitative picture of landscape SUSY was 
presented. Probability distributions for Higgs and sparticles masses
were derived from landscape considerations in Ref.~\cite{land} while
in Ref. \cite{land3}, predictions for landscape SUSY were compared to LHC
simplified model searches along with WIMP direct and indirect detection searches. In Ref. \cite{Baer:2019uom}, similar methods were applied to determination 
of the Peccei-Quinn scale in SUSY axion models.
Some complementary investigations on the likelihood of $N=1$ 
SUSY emerging from the heterotic landscape have been performed in 
Ref. \cite{Dienes:2007ms}.

\section{Prior distributions and selection criteria for landscape SUSY}
\label{sec:lsusy}

A simple ansatz for the distribution of string vacua in terms of 
SUSY breaking scales $m_{hidden}^2$ (where the soft breaking mass scale 
$m_{soft}=m_{hidden}^2/m_P$ and $m_P$ is the reduced Planck mass) is given by 
\be
dN_{vac}[m_{hidden}^2,m_{weak},\Lambda_{cc}]= 
f_{SUSY}(m_{hidden}^2)\cdot f_{EWFT}\cdot f_{cc}\cdot dm_{hidden}^2 .
\label{eq:dNvac}
\ee
The cosmological constant finetuning penalty is expected to be
$f_{cc}\sim \Lambda_{cc}/m^4$ where initial expectations were that 
$m^4$ was taken to be $m_{hidden}^4$.
In the 4-d supergravity effective theory which emerges after string compactification, 
the cosmological constant is given by
\be
\Lambda_{cc} \,\, = \,\, m_{hidden}^4\, - \, 3 e^{K/m_P^2}|W|^2/m_P^2 \,\,
\ee
where $m_{hidden}^4=\sum_i |F_i|^2 \, +  \, \frac{1}{2}\sum_\alpha D^2_\alpha $ 
is a mass scale associated with the hidden sector 
(and usually in SUGRA-mediated models it is assumed $m_{hidden}\sim 10^{12}$ GeV 
such that the gravitino gets a mass $m_{3/2}\sim m_{hidden}^2/m_P$).

A key observation of Susskind\cite{suss} and Denef and Douglas\cite{denefdouglas} (DD)
was that $W$ at the minima is distributed uniformly as a complex variable, 
and the distribution of $e^{K/m_P^2}|W|^2/m_P^2$
is not correlated with the distributions of $F_i$ and $D_\alpha$.
Setting the cosmological constant to nearly zero, then, has no effect on the distribution of
supersymmetry breaking scales.
Physically, this can be understood by the fact that the superpotential 
receives contributions from many sectors of the theory, 
supersymmetric as well as non-supersymmetric. 
In this case, the $m^4$ in $f_{cc}$ should be taken to be $m_{string}^4$ instead of
$m_{hidden}^4$, rendering this term inconsequential to how the number of vacua
are distributed in terms of $m_{soft}$.

Another key observation from examining flux vacua in IIB string theory
is that the SUSY breaking $F_i$ and $D_\alpha$ terms
are likely to be uniformly distributed-- 
in the former case as complex numbers while in the latter case as real numbers.
Then one expects the following distribution of supersymmetry breaking scales
\be
f_{SUSY}(m_{hidden}^2) \, \sim \, (m^2_{hidden})^{2n_F+n_D - 1}
\label{eq:fSUSY}
\ee
where $n_F$ is the number of $F$-breaking fields and $n_D$ is the number of 
$D$-breaking fields in the hidden sector.
For just a single $F$-breaking term, then one expects a {\it linear} 
statistical draw towards large soft terms $f_{SUSY}\sim m_{soft}^n$
where $n=2n_F+n_D-1$ and in this case $n=1$.
For SUSY breaking contributions from multiple hidden sectors, as typically
expected in string theory, then $n$ can be much larger, with a consequent 
stronger pull towards large soft breaking terms.

An initial guess for $f_{EWFT}$, the (anthropic) finetuning factor,
was $m_{weak}^2/m_{soft}^2$ which would penalize soft terms which were much
bigger than the weak scale.
This ansatz fails on several points.
\begin{itemize}
\item Many soft SUSY breaking choices will land one into charge-or-color breaking (CCB) minima of the EW scalar potential. Such vacua would likely not lead to a livable
universe and should be vetoed.
\item Other choices for soft terms may not even lead to EW symmetry breaking (EWSB).
For instance, if $m_{H_u}^2(\Lambda )$ is too large, 
then it will not be driven negative to trigger spontaneous EWSB. 
These possibilities also should be vetoed.
\item In the event of appropriate EWSB minima, then sometimes {\it larger} 
high scale soft terms lead to {\it more natural} weak scale soft terms. 
For instance, if $m_{H_u}^2(\Lambda )$ is
large enough that EWSB is {\it barely broken}, then $|m_{H_u}^2(weak)|\sim m_{weak}^2$.
Likewise, if the trilinear soft breaking term $A_t$ is big enough, then there 
is large top squark mixing and the $\Sigma_u^u(\tst_{1,2})$ terms enjoy large
cancellations, rendering them $\sim m_{weak}^2$. 
The same large $A_t$ values lift the Higgs mass $m_h$ up to the 125 GeV regime.
\end{itemize}

Here, we will assume a {\it natural} solution to the SUSY $\mu$ problem\cite{mupaper}.
A recent possibility is the hybrid CCK or SPM models\cite{Z24R} 
which are based on a ${\mathbb Z}_{24}^R$ discrete $R$ symmetry which can emerge 
from compactification of extra dimensions in string theory. 
The ${\mathbb Z}_{24}^R$ symmetry is strong enough to allow a gravity-safe $U(1)_{PQ}$ 
symmetry to emerge (which solves the strong CP problem) while also forbidding RPV terms
(so that WIMP dark matter is generated). 
Thus, both Peccei-Quinn (PQ) and $R$-parity conservation (RPC) arise as approximate accidental
symmetries similar to the way baryon and lepton number conservation emerge accidentally
(and likely approximately) due to the SM gauge symmetries. 
These hybrid models also solve the SUSY $\mu$ problem via a 
Kim-Nilles\cite{KN} operator so that 
$\mu\sim \lambda_\mu f_a^2/m_P$ and $\mu\sim 100-200$ GeV (natural) for $f_a\sim 10^{11}$ GeV
(the sweet zone for axion dark matter). 
The ${\mathbb Z}_{24}^R$ symmetry also suppresses dimension-5 proton decay operators\cite{lrrrssv2}. 

\begin{figure}[!htbp]
\begin{center}
\includegraphics[height=0.75\textheight]{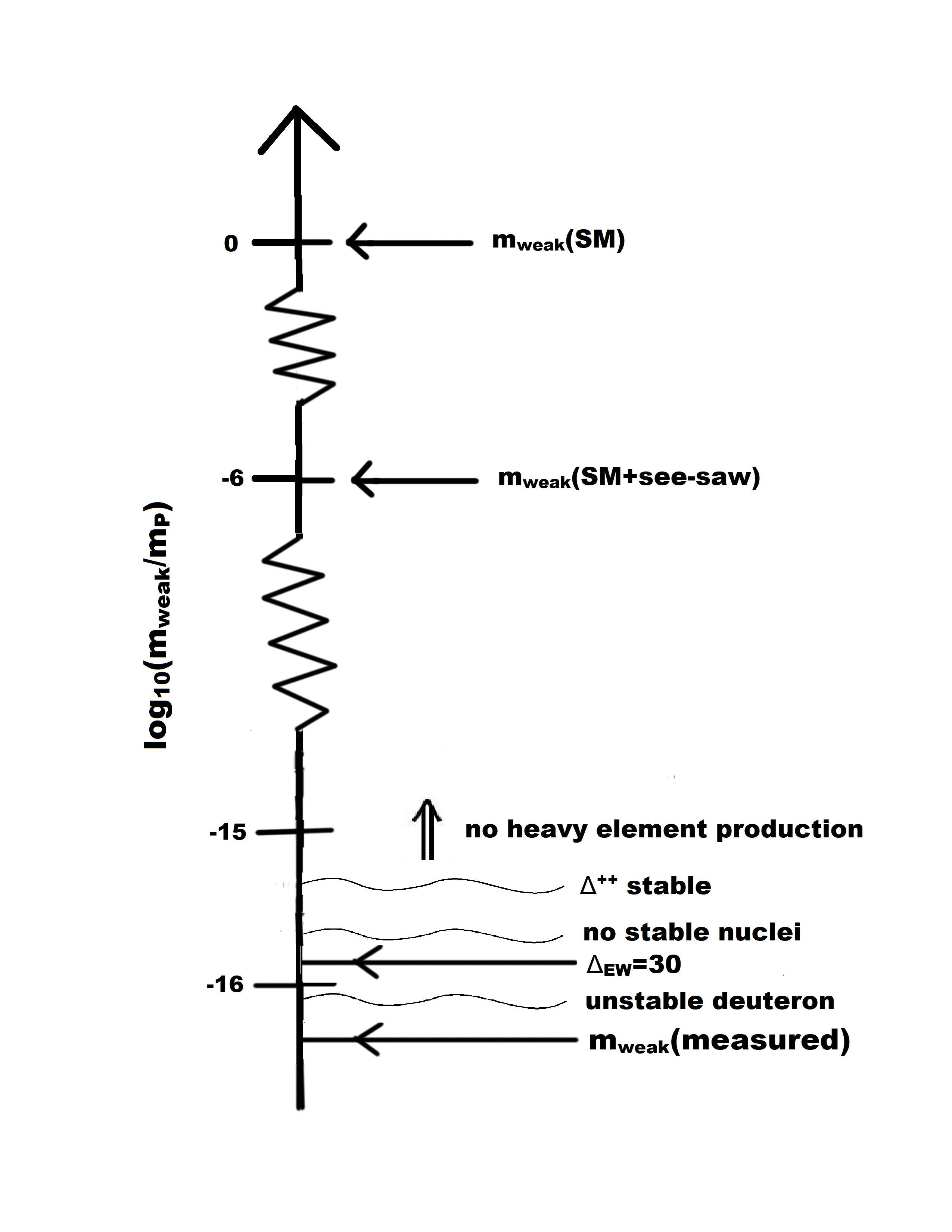}
\caption{Value of $m_{weak}$ as predicted when the SM is valid up to
energy scale 1.) $Q=m_P$, 2.) the neutrino see-saw scale and 
3.) valid just up to the measured value of $m_{weak}\sim 100$ GeV.
We also show several anthropic bounds on $m_{weak}$ from nuclear physics
considerations of Agrawal {\it et al.}\cite{Agrawal:1997gf}.
\label{fig:mweak}}
\end{center}
\end{figure}

Once a natural value of $\mu\sim 100-300$ GeV is obtained, then we may invert the usual usage
of Eq. \ref{eq:mzs} to determine the value of the weak scale in various pocket 
universes (with MSSM as low energy effective theory) for a given choice of soft terms.
Based on nuclear physics calculations by Agrawal {\it et al.}\cite{Agrawal:1997gf}, a pocket universe value of $m_{weak}^{PU}$ which deviates from our measured value by a factor 2-5 is likely to lead
to an unlivable universe as we understand it. Weak interactions and fusion processes 
would be highly suppressed and even complex nuclei could not form. The situation is 
shown in Fig. \ref{fig:mweak}. 
We will adopt a conservative value that the weak scale should not
deviate by more than a factor four from its measured value. This corresponds to a 
value of $\Delta_{EW}\alt 30$.
Thus, for our final form of $f_{EWFT}$ we will adopt $f_{EWFT}=\Theta(30-\Delta_{EW})$
while also vetoing CCB or no EWSB vacua.
\newpage
\section{Why the SM is likely a rare occurance in the landscape}
\label{sec:sm}

\begin{figure}[!htbp]
\begin{center}
\includegraphics[height=0.35\textheight]{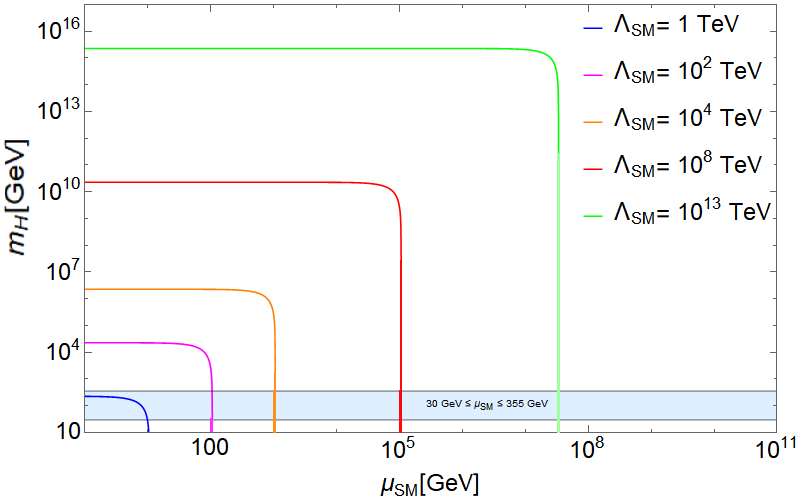}
\caption{The value of the SM Higgs mass $m_H$ versus SM $\mu_{SM}$ parameter 
for theory cut-off values $\Lambda_{SM}=1$, $10^2$, $10^4$, $10^{8}$ and 
$10^{13}$ TeV. The anthropic band is shown in blue.
\label{fig:mhSM}}
\end{center}
\end{figure}

Before using Eq. \ref{eq:dNvac} to evaluate various SUSY models, 
we will first examine the case of EWSB in the SM. 
For the SM with
\be
V_{SM}=-\mu_{SM}^2\phi^\dagger\phi+\lambda_{SM} (\phi^\dagger\phi)^2 ,
\ee
then the Higgs mass, including quadratic divergent radiative corrections, is found to be
\be
m_H^2\simeq m_H^2(tree)+\delta m_H^2
\ee
where $m_H^2(tree)=2\mu_{SM}^2$ and $\delta m_H^2=\frac{3}{4\pi^2}\left(-\lambda_t^2+\frac{g^2}{4}+
\frac{g^2}{8\cos^2\theta_W}+\lambda_{SM}\right)\Lambda_{SM}^2$,
and where $\Lambda_{SM}$ is the mass scale cut-off beyond which the SM is no longer the appropriate low energy effective field theory.
To gain the measured value of $m_H\simeq 125$ GeV, then for $\Lambda_{SM}\gg m_H$
we can freely tune $m_{H}^2(tree)$ to compensate for the large radiative corrections.
The situation is shown in Fig. \ref{fig:mhSM} where we show the required value of
$\mu_{SM}$ needed to gain $m_{H}=125$ GeV for various choices of $\Lambda_{SM}$. 
Since nothing in the SM favors any particular value of $\mu_{SM}$, we will assume its value
is uniformly distributed (logarithmically over the decades of values) in the landscape.
It is plain to see that for $\Lambda_{SM}=1$ TeV, then a wide range of values for
$\mu_{SM}$ leads to a weak scale (typified here by $m_{H}$) within the Agrawal band of allowed
values. However, for $\Lambda_{SM}\gg m_{weak}$, then only a tiny (finetuned) range of
$\mu_{SM}$ values leads to a viable value for $m_{weak}$. In this case, stringy naturalness
and conventional natural coincide in that an anthropically allowed value of the
weak scale requires that the SM be a valid effective field theory only for cut-off
value $\Lambda_{SM}\alt 1$ TeV.
\newpage
\section{Why CMSSM/mSUGRA is likely an infrequent occurance in the
landscape}
\label{sec:cmssm}

In SUSY models, all the quadratic divergences cancel, leaving only log divergences
whose effects may be captured via renormalization group (RG) equations.
Thus, SUSY models carry with them a solution to the Big Hierarchy problem.
The question  then is: do they carry with them a Little Hierarchy problem?

The CMSSM or mSUGRA model\cite{cmssm} is defined by GUT scale input parameters
\be
m_0,\ m_{1/2},\ A_0,\ \tan\beta,\ sign(\mu)\ \ \ \ (CMSSM/mSUGRA)
\ee
where $m_0$ is the GUT scale unified scalar mass where 
$m_{H_u}=m_{H_d}=m_0$, $m_{1/2}$ is the unified gaugino mass, $A_0$ is 
the unified trilinear soft breaking term and $sign(\mu )=\pm 1$.
The CMSSM has served for many years as a sort of SUSY paradigm model
for SUSY collider and dark matter signal predictions. In CMSSM, the weak scale soft terms
are derived from RG running of the unified soft terms from $Q=m_{GUT}$ to $Q=m_{weak}$,
where then mixings and mass eigenstates may be evaluated. In the CMSSM model, 
since $m_{H_u}^2$ is input at the $m_{GUT}$ scale, then its weak scale value is
derived. The $\mu$ term is (fine)tuned via Eq. \ref{eq:mzs} to gain a value in accord
with the measured $Z$-boson mass.\footnote{We compute SUSY spectra in all models using
Isajet 7.88\cite{isajet}.}

In years past, it was possible to find regions of CMSSM parameter space with small
$\mu$ as required for naturalness in the hyperbolic branch\cite{ccn} 
or focus point region\cite{fmm} (HB/FP). The HB/FP region can appear for $m_0\alt 10$ TeV
for small values of $A_0\sim 0$. However, the measured value of $m_h\simeq 125$ GeV
requires large $A_t$ parameter which then moves the HB/FP region out to huge $m_0$
values where the $\Sigma_u^u(\tst_{1,2})$ are large, rendering the model unnatural.
Detailed scans of the CMSSM model parameter space in accord with requiring 
$m_h=125\pm 2$ GeV find that the minimal value of $\Delta_{EW}$ is around 100 
but where typically $\Delta_{EW}$ can range up to $10^4$\cite{seige}.

To gain some insight on how frequent models with large values of $\Delta_{EW}$ might
occur in the landscape, we take the limit of Eq. \ref{eq:mzs} wherein the radiative corrections
are small so that $m_Z^2\simeq -2m_{H_u}^2-2\mu^2$ and then consider SUSY models
where $m_{H_u}^2$ is driven to large negative values at the weak scale. 
We can replace $-m_{H_u}^2(weak)$ by $\Delta_{EW}\cdot m_Z^2(measured)/2$
to use Eq. \ref{eq:mzs} to determine the pocket-universe (PU) value of $m_Z^{PU}$
which is expected in SUSY models within the landscape in terms of $\Delta_{EW}$ and $\mu$.

\begin{figure}[!htbp]
\begin{center}
\includegraphics[height=0.4\textheight]{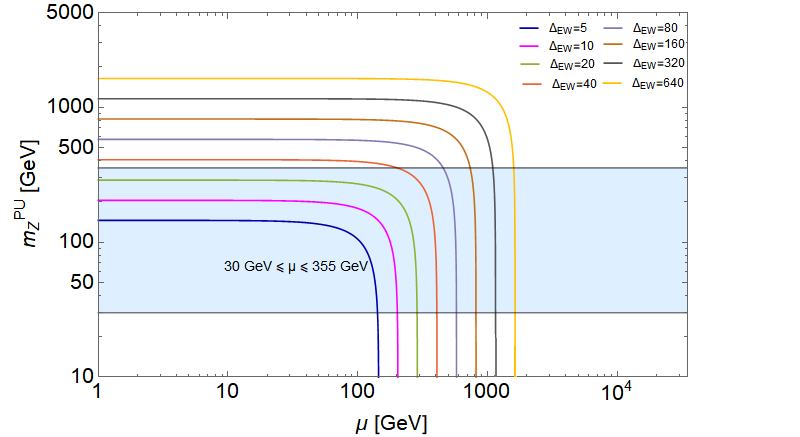}
\caption{The pocket universe value of $m_Z^{PU}$  versus the SUSY  $\mu$ 
parameter for various values of EW finetuning parameter $\Delta_{EW}$.
The anthropic band is shown in blue.
\label{fig:mzPU}}
\end{center}
\end{figure}

In Fig.~\ref{fig:mzPU} we plot the value of $m_Z^{PU}$ versus $\mu$ for a variety of
choices of $\Delta_{EW}$ ranging from natural values $\sim 5-20$ up to as large as 640.
We also show the shaded band in Fig. \ref{fig:mzPU} which corresponds to 
values of $m_Z^{PU}$ in accord with the Agrawal {\it et al.} allowed values
which should give rise to a habitable pocket universe.
Assuming a uniform distribution of $\mu$ values in the landscape, then we 
see from the figure that for low, natural values of $\Delta_{EW}$ there are significant
ranges of $\mu$ which lead to values of $m_Z^{PU}$ within the anthropic zone. However, as
$\Delta_{EW}$ increases, the expected value of $m_Z^{PU}$ increases well beyond the 
anthropic allowed zone unless one tunes $\mu$ to lie within a tightening range 
of (finetuned) values. Thus, we would expect that SUSY models such as CMSSM/mSUGRA-- 
where $\Delta_{EW}$ cannot attain natural values for $m_h\sim 125$ GeV-- to be rather
infrequent occurances of our fertile patch of the landscape which contains the MSSM as the
low energy effective theory. 

\section{Radiative natural SUSY from stringy naturalness}
\label{sec:rns}

From Fig. \ref{fig:mzPU}, we see that models with a weak scale value of the $\mu$ parameter
hold a higher likelihood of landing within the narrow band of allowed (pocket universe) 
weak scale values from Agrawal {\it et al.}\cite{Agrawal:1997gf}. 
Models with non-universal Higgs masses where $m_0\ne m_{H_u}$ such as
the two-extra-parameter non-universal Higgs model\cite{nuhm2} (NUHM2) or its extension for
non-universal generations NUHM3 (where $m_0(1,2)\ne m_0(3)$ as suggested by mini-landscape 
models\cite{miniland}) are applicable in this situation since the high scale Higgs masses
$m_{H_u}^2$ and $m_{H_d}^2$ can be traded for weak scale inputs $\mu$ and $m_A$. Thus, 
the NUHM2 model has input parameters $m_0,\ m_{1/2},\ A_0,\ \tan\beta,\ \mu,\ m_A$.
The added Higgs mass non-universality (which is expected since the Higgs multiplets live in 
different GUT multiplets from matter scalars) allows for small $\mu\sim 100-300$ GeV
for any points in model parameter space.

\subsection{Living dangerously}
\label{ssec:LD}

\begin{figure}[tbp]
\begin{center}
\includegraphics[height=0.39\textheight]{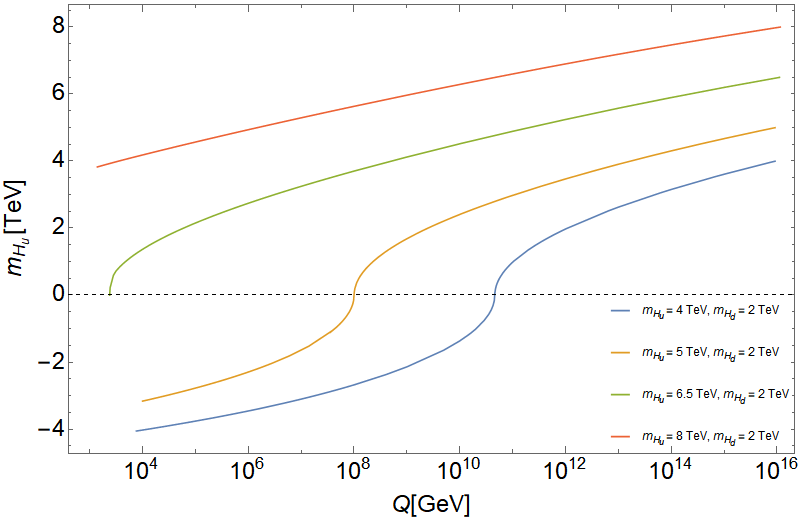}
\caption{Running of $m_{H_u}^2$ vs. $Q$ for several choices of
$m_{H_u}^2(\Lambda )$ in the NUHM2 model for $m_0=5$ TeV, 
$m_{1/2}=1.2$ TeV, $A_0=-8$ TeV and $\tan\beta =10$ for values
of $m_{H_{u,d}}$ shown in the figure. 
The radiatively-driven natural SUSY (RNS) case is green. 
We also show several unnatural SUSY model parameters.
\label{fig:mHu2Q}}
\end{center}
\end{figure}

In Fig. \ref{fig:mHu2Q}, we show the running of the up-Higgs soft mass (actually
$sign(m_{H_u}^2)\cdot\sqrt{|m_{H_u}^2|}$) in the NUHM2 model for parameters
$m_0=5$ TeV,  $m_{1/2}=1.2$ TeV, $A_0=-8$ TeV and $\tan\beta =10$ with $m_{H_d}=2$ TeV but for
$m_{H_u}(\Lambda )=4,\ 5,\ 6.5$ and 8 TeV. We see that for the lower values of $m_{H_u}(\Lambda )$, 
then $m_{H_u}^2$ runs deeply negative leading to a large negative weak scale value
of $m_{H_u}^2(weak)$. As $m_{H_u}^2(\Lambda )$ increases, then it is driven to smaller
and smaller (more natural) weak scale values. For too large a value of $m_{H_u}^2(\Lambda )$, 
then its running value isn't driven negative at the weak scale and radiative EWSB does not occur.
Such unphysical pocket universe solutions must be vetoed. Thus, we see that the
landscape pull on the soft SUSY breaking term $m_{H_u}^2(\Lambda )$ to large values is just what
is needed to gain a natural value of $m_{H_u}^2$ at the weak scale. 
This sort of situation has been dubbed as {\it living dangerously} 
in Ref's \cite{ArkaniHamed:2005yv} and \cite{Giudice:2006sn} since the soft term 
is selected to be as large as possible such that EW symmetry is barely broken.\footnote{
Arkani-Hamed and Dimopoulos\cite{ArkaniHamed:2005yv}  state:
``anthropic reasoning leads to the conclusion that we live dangerously close to
violating an important but fragile feature of the low-energy world...'', in this
case, appropriate electroweak symmetry breaking.}
In Ref's \cite{ltr,rns}, this situation is called radiatively-driven naturalness, or
{\it radiative natural SUSY}, since the
largest viable $m_{H_u}^2(\Lambda )$ soft terms lead to natural values of $m_{H_u}^2$ at the weak scale.

\begin{figure}[htbp]
\begin{center}
\includegraphics[height=0.3\textheight]{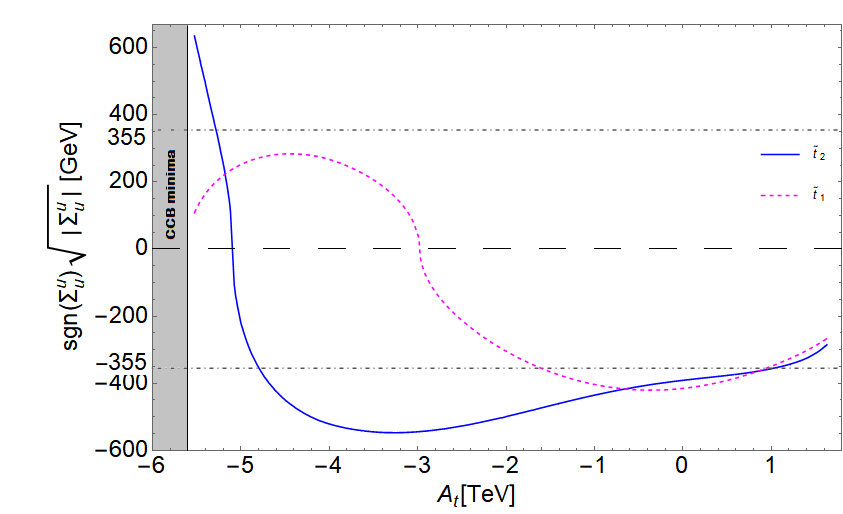}
\caption{Contributions $sign (\Sigma_u^u)\sqrt{|\Sigma_u^u(\tst_{1,2})|}$ 
to the weak scale vs. $A_t(weak)$ in the NUHM2 model with $m_0=5$ TeV, 
$m_{1/2}=1.2$ TeV, $m_A=2$ TeV, $\mu =200$ GeV and $\tan\beta =10$.
\label{fig:Sigt1t2}}
\end{center}
\end{figure}

A second example of living dangerously within the string landscape is shown in
Fig.~\ref{fig:Sigt1t2}, where we show the contributions $\Sigma_u^u(\tst_{1,2})$ 
to the weak scale vs. $A_t$ for the same NUHM2 parameter choices as in Fig. \ref{fig:mHu2Q}.
Here, we see the contribution $\Sigma_u^u(\tst_{1,2})$ are rather large negative for
$A_0\sim 0$ with $sign(\Sigma_u^u(\tst_{1,2}))\cdot\sqrt{|\Sigma_u^u(\tst_{1,2})|}\sim -400$ GeV
(in which case we might expect a pocket universe weak scale of $m_Z^{PU}\sim 400$ GeV).
As $A_0$ moves to large negative values, then we find a point around $A_0\sim -5$ TeV where
both terms become small, yielding contributions to  the weak scale which are indeed 
comparable to our universe's measured value. For much larger (negative) values of $A_0$,
then we rapidly move into the zone of charge-and-color-breaking (CCB) minima since
top squark squared-mass soft terms are driven negative. Thus, in this case, the
$A_0$ trilinear soft term is statistically preferred to be large (negative) values--
but stopping short of CCB minima. 
This again leads to natural contributions to the weak scale.

\begin{figure}[!htbp]
\begin{center}
\includegraphics[height=0.35\textheight]{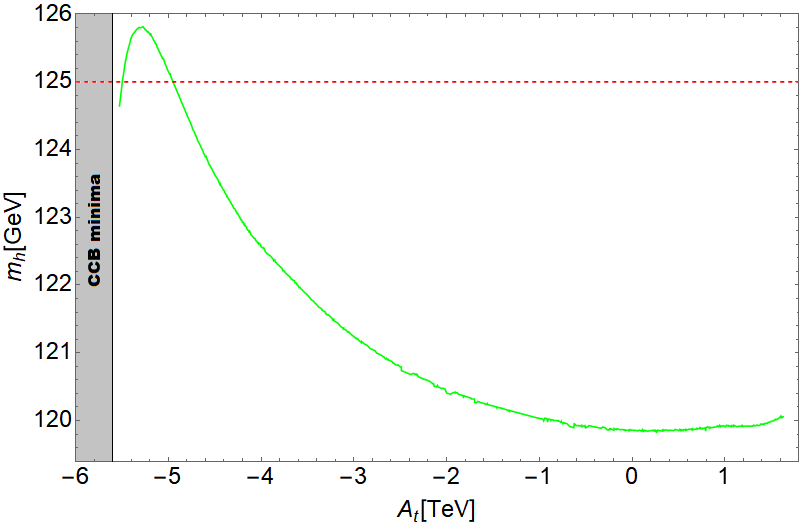}
\caption{Value of $m_h$ vs. $A_t$ in the NUHM2 model with $m_0=5$ TeV, 
$m_{1/2}=1.2$ TeV, $m_A=2$ TeV, $\mu =200$ GeV and $\tan\beta =10$.
\label{fig:mh}}
\end{center}
\end{figure}

\begin{figure}[!htbp]
\begin{center}
\includegraphics[height=0.21\textheight]{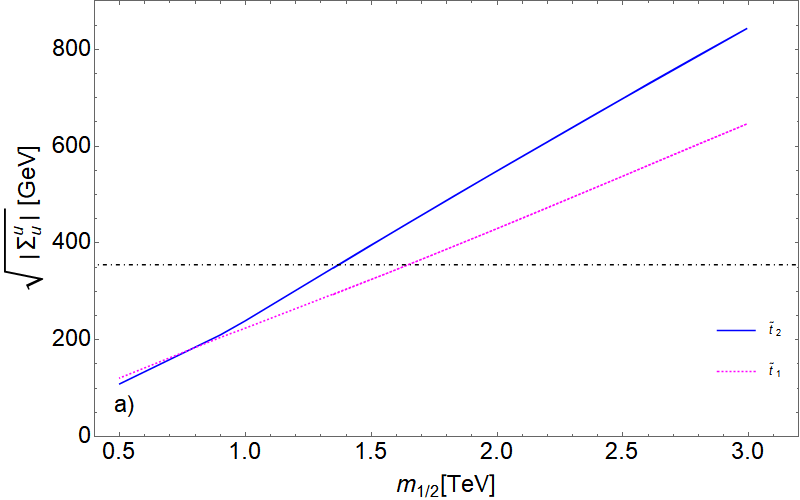}
\includegraphics[height=0.22\textheight]{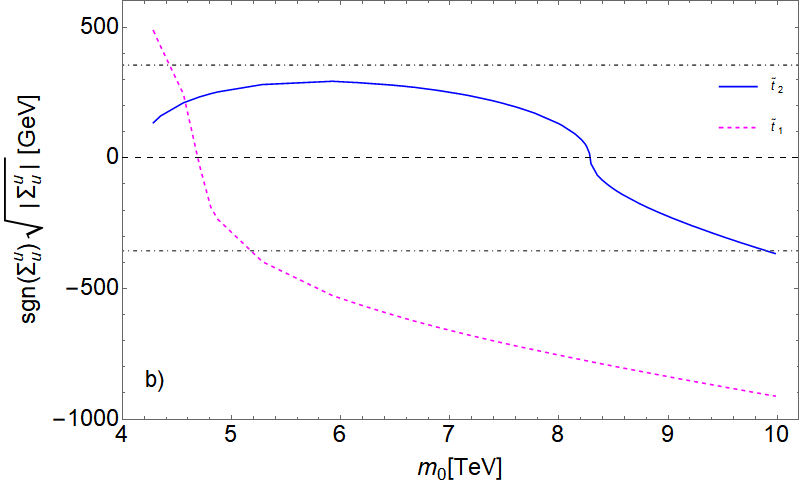}
\caption{Plot of $\sqrt{|\Sigma_u^u(\tst_{1,2})|}$ vs. {\it a}) $m_{1/2}$ 
for $m_0=5$ TeV and {\it b}) $m_0$ for $m_{1/2}=1.2$ TeV in the NUHM2 model.
We also take $A_0=-8$ TeV, $\mu =200$ GeV, $m_A=2$ TeV and $\tan\beta =10$.
The horizontal lines shows where contributions to $m_{weak}$ exceed a factor four
times its measured value in our pocket universe.
\label{fig:sigmavsmhf_m0}}
\end{center}
\end{figure}

A beautiful consequence of the statistical draw to large (negative) $A_0$ values is 
that this induces large mixing in the top-squark sector, which also ends up
maximizing the value of $m_h$\cite{mhiggs,h125}. 
The case here is shown in Fig. \ref{fig:mh} where we plot the light Higgs mass 
$m_h$ versus $A_0$ for the same parameters as in Fig. \ref{fig:Sigt1t2}. 
For an unnatural value of $A_0\sim 0$, then we expect $m_h\sim 119$ GeV. However, as
$A_0$ increases to large (negative) values, then $m_h\to 124-126$ GeV, in accord with
its measured value in our universe.
\newpage
A third example occurs if we allow the landscape to statistically select large
values of $m_{1/2}$. This is shown in Fig. \ref{fig:sigmavsmhf_m0}{\it a})
where we plot $\sqrt{|\Sigma_u^u(\tst_{1,2})|}$ vs. $m_{1/2}$ for
fixed $m_0=5$ TeV and other parameters fixed as in the Caption. 
In this case, then as the gaugino masses increase,
especially the gluino mass, then large $SU(3)_C$ contributions $M_3$ to the
top-squark soft masses tend to drive them to large values, thus increasing
the $\Sigma_u^u(\tst_{1,2})$ contributions to values well beyond our measured
value of the weak scale. In such cases, we would expect too large a value of
the weak scale, typically in violation of Agrawal {\it et al.} bounds. 
These would lead to violations of the so-called ``atomic principle'', and atoms
as we know them would not form with too large a value of $m_{weak}$.
In Fig. \ref{fig:sigmavsmhf_m0}{\it b}), we show the values of
$\Sigma_u^u(\tst_{1,2})$ vs. $m_0$ for fixed $m_{1/2}=1.2$ TeV. In this case, if
the scalar soft terms become too large, then again the values
of $\sqrt{|\Sigma_u^u(\tst_{1,2})|}$ become too large and we would gain a
pocket universe weak scale value in excess of bounds from nuclear/atomic
anthropic requirements.

\subsection{Naturalness vs. stringy naturalness}
\label{ssec:NvSN}

\begin{figure}[tbp]
\begin{center}
\includegraphics[height=0.24\textheight]{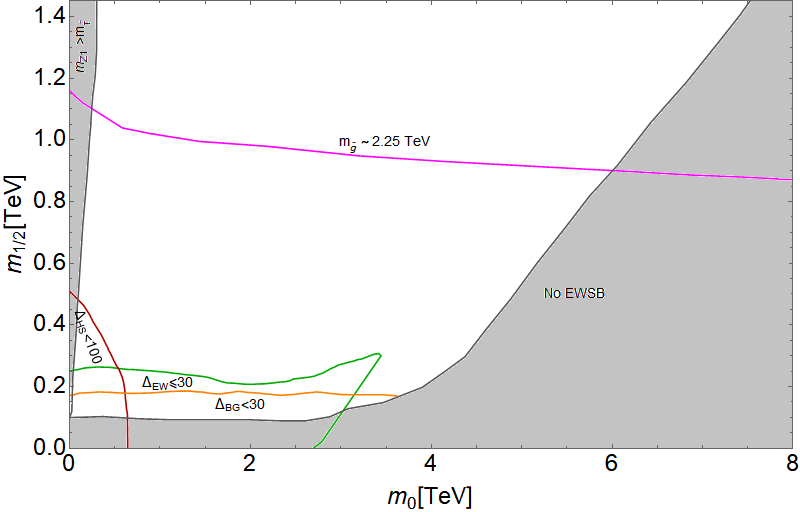}
\includegraphics[height=0.24\textheight]{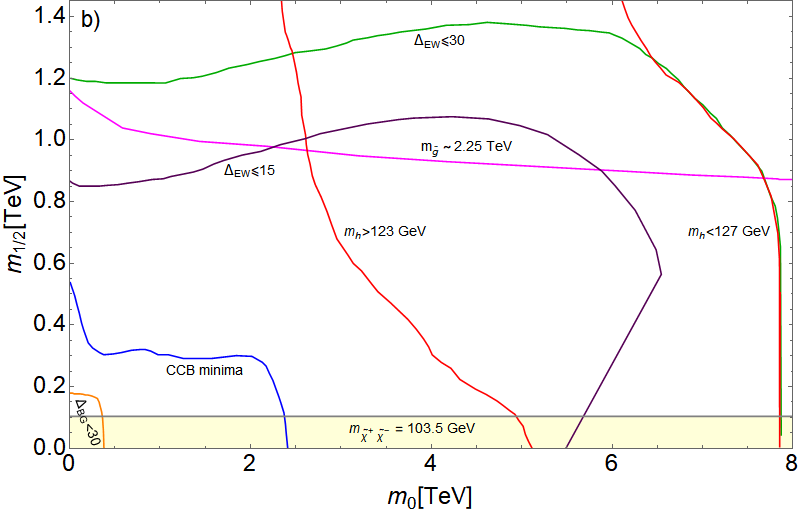}
\caption{The $m_0$ vs. $m_{1/2}$ plane of {\it a}) the mSUGRA/CMSSM model 
with $A_0=0$ and {\it b}) the NUHM2 model with $A_0=-1.6 m_0$, $\mu =200$ GeV 
and $m_A=2$ TeV. For both cases, we take $\tan\beta =10$.
We show contours of various finetuning measures along with Higgs mass 
contours and LEP2 and LHC Run 2 search limits.
\label{fig:m0mhf}}
\end{center}
\end{figure}

Next, we would like to explore  how the notion of stringy naturalness
compares to conventional naturalness measures.
To begin, we plot naturalness contours in Fig. \ref{fig:m0mhf}{\it a}) 
in the $m_0$ vs. $m_{1/2}$ plane of the CMSSM/mSUGRA model for $A_0=0$, $\tan\beta =10$
and $\mu >0$. The $\Delta_{HS}<100$ contour shows the region of lighter top squarks
for small $A_0$ parameter in the low $m_0$, low $m_{1/2}$ region.
The orange contour denotes where $\Delta_{BG}<30$ while the green contour
denotes where $\Delta_{EW}<30$. These latter two measure reflect focus point behavior
in that TeV-scale stops are still natural (since the contours are relatively flat 
with variation in $m_0$). 
The LHC limit on $m_{\tg}>2.25$ TeV is shown as the magenta contour. 
This plane might lead to skepticism regarding weak scale SUSY since the LHC 
allowed region is so far beyond the naturalness upper bounds. 
Also, in this plane the light Higgs mass $m_h$ is always below 123 GeV.
The important lesson for now is that the {\it more} natural regions occur
at the lowest $m_0$ and $m_{1/2}$ values, where the various measures are smallest, 
and the sparticle masses are closest to the measured weak scale.

For comparison, we show the same $m_0$ vs. $m_{1/2}$ plane in Fig. \ref{fig:m0mhf}{\it b}), 
but this time for the NUHM2 model where $\mu =200$ GeV, $m_A=2$ TeV, 
$A_0=-1.6 m_0$ and as before $\tan\beta =10$. In this case, the lower left region of the
plot actually leads to CCB minima (blue contour) so that $\Delta_{HS}$ is not computable.
$\Delta_{BG}$ is still computable and has shrunk down into the lower-left due to large
$A_0$ term contributions. However, we now see that the $\Delta_{EW}$ values have expanded
out to much larger $m_0$ and $m_{1/2}$ values since it is largely determined by the
$\Sigma_u^u(\tst_{1,2})$ values, since $\mu$ is fixed to be near the measured EW scale.
Here, a substantial amount of natural SUSY parameter space lies beyond LHC gluino mass
limits. Nonetheless, the lower portions of $m_0$, $m_{1/2}$ space are more natural
since they yield smaller values of $\Delta_{EW}$. 
Also, we show contours of $m_h=123$ and 127 GeV. The region with $m_h=125\pm 2$ GeV 
overlaps nicely with the natural SUSY region, with plenty of parameter space 
beyond the LHC gluino mass limit.  

\begin{figure}[!htbp]
\begin{center}
\includegraphics[height=0.24\textheight]{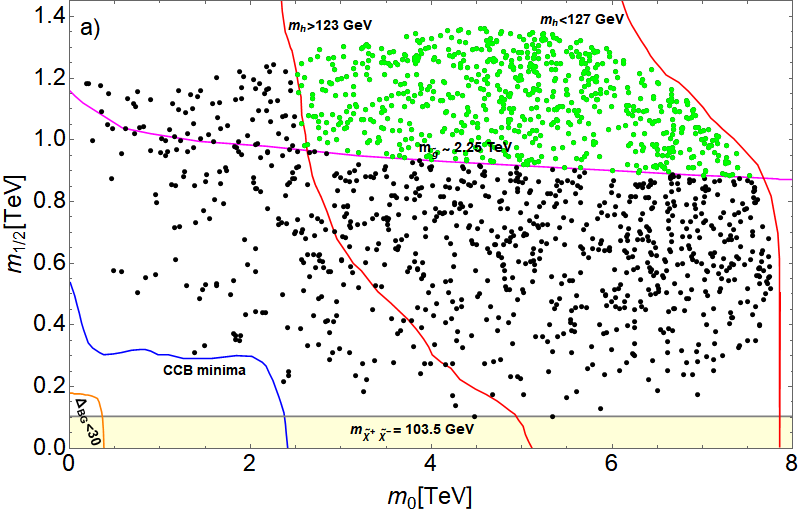}
\includegraphics[height=0.24\textheight]{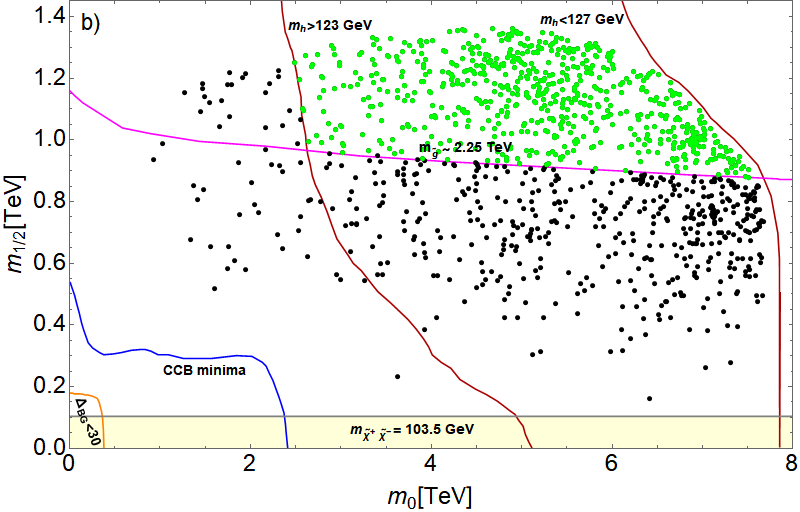}\\
\includegraphics[height=0.24\textheight]{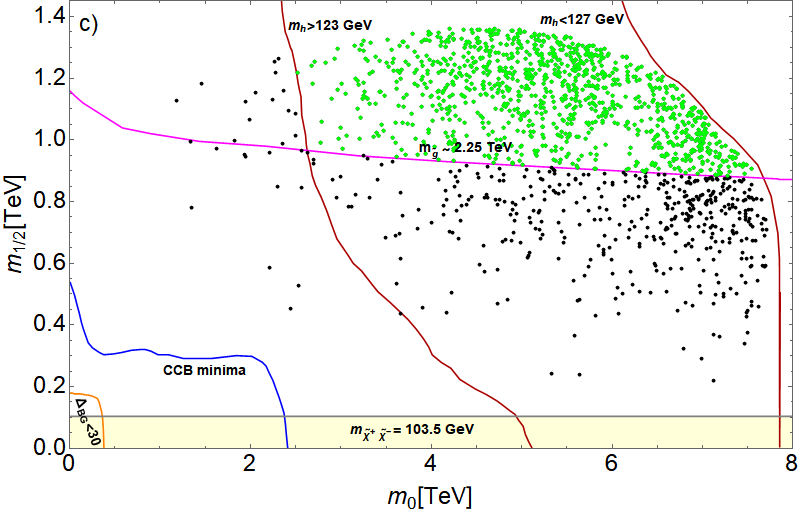}
\includegraphics[height=0.24\textheight]{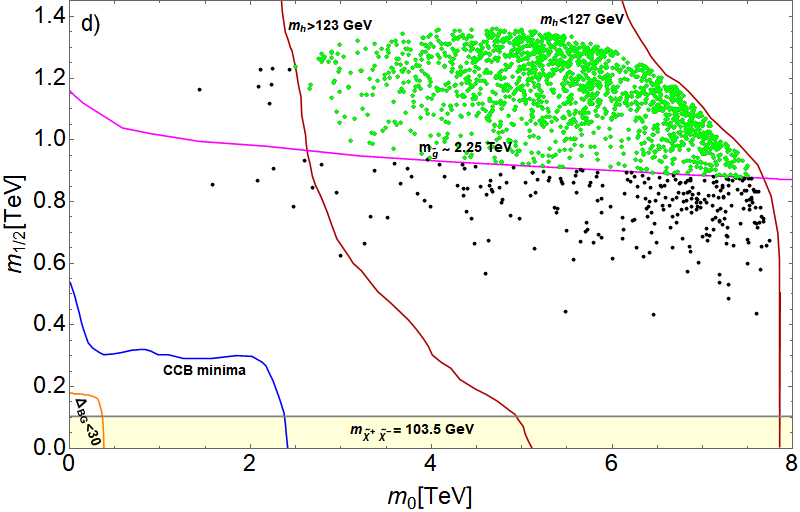}
\caption{The $m_0$ vs. $m_{1/2}$ plane of the NUHM2 model 
with $A_0=-1.6 m_0$, $\mu =200$ GeV and $m_A=2$ TeV and
{\it a}) an $n=1$ draw on soft terms,
{\it b}) an $n=2$ draw, {\it c}) an $n=3$ draw and {\it d}) an $n=4$ draw.
The higher density of points denotes greater stringy naturalness.
The LHC Run 2 limit on $m_{\tg}>2.25$ TeV is shown by the magenta curve.
The lower yellow band is excluded by LEP2 chargino pair search limits.
\label{fig:m0mhfn}}
\end{center}
\end{figure}

Now we would like to compare the previous natural SUSY regions against the
regions preferred by Douglas' stringy naturalness. To accomplish this, next we
show again in Fig's \ref{fig:m0mhfn}{\it a-d}) 
the $m_0$ vs. $m_{1/2}$ plane for the NUHM2 model with the same
parameters as in Fig.~\ref{fig:m0mhf}{\it b}). We generate SUSY soft parameters
in accord with Eq. \ref{eq:dNvac} for various values of $n=2n_F+n_D-1=1,\ 2,\ 3$ 
and 4.\footnote{The high $n$ values allow for a consistent sampling of NUHM2 parameter space
since here we fix the $A_0$ parameter in terms of $m_0$ so it never gets too large (which
would lead to CCB minima and non-anthropic vacua) as compared with Ref. \cite{land}.}
The more stringy natural regions of parameter space are denoted by the higher
density of sampled points.

In Fig. \ref{fig:m0mhfn}{\it a}), we show the stringy natural regions for the case
of $n=1$. Of course, no dots lie below the CCB boundary since such minima must be vetoed
as they likely lead to an unlivable pocket universe. Beyond the CCB contour, the solutions
are in accord with livable vacua. 
But now the density of points {\it increases} with increasing $m_0$ and $m_{1/2}$ 
(linearly, for $n=1$), showing that the more stringy natural regions lie at the 
{\it highest} $m_0$ and $m_{1/2}$ values which are consistent with 
generating a weak scale within the Agrawal bounds. 
Beyond these bounds, the density of points of course drops to zero 
since contributions to the weak scale exceed its measured value by a factor 4. 
There is some fluidity of this latter bound as indicated in Fig. \ref{fig:mweak},
so values of $\Delta_{EW}\sim 20-40$ might also be entertained 
The result that stringy naturalness for
$n\ge 1$ favors the largest soft terms (subject to $m_Z^{PU}$ not ranging too far from
our measured value) stands in stark contrast to conventional naturalness
which favors instead the lower values of soft terms. Needless to say, the stringy natural
favored region of parameter space is in close accord with LHC results in that
LHC find $m_h=125$ GeV with no sign yet of sparticles.

In Fig's \ref{fig:m0mhfn}{\it b-d} we show the same $m_0$ vs. $m_{1/2}$ planes but
for $n=2$, 3 and 4. As $n$ steadily increases, the stringy natural region
is pushed more strongly to large values of $m_0$ and $m_{1/2}$ so that relatively few
vacua lie below the LHC gluino mass limit. Indeed, we would say that the 
stringy naturalness prediction is that LHC should see a Higgs mass around 125 GeV with no sign
yet of sparticles.

\section{Why high scale SUSY is likely a rare occurance in the landscape}
\label{sec:hsSUSY}

While it is often argued that the landscape opens new territory in model building
and motivation for finetuned SUSY models, here we will argue that such models
should be, while possible, relatively rare occurances on the string theory landscape.
The rationale is usually that since the cosmological constant is finetuned, 
then perhaps a similar mechanism allows for a finetuned weak scale.
Here we would counter with two observations. 
First, in Weinberg's approach, the cosmological constant is about 
as natural as possible subject to allowing for pocket universes that allow for
galaxy condensation. Second, there is at present no plausible alternative for
a tiny cosmological constant other than the landscape selection. As alternatives
to various high scale SUSY models described below, SUSY with radiatively driven naturalness
should be a relatively common occurance on the landscape, as argued above, whereas
finetuned models, while possible, should be relatively rare.

\begin{itemize}
\item {\it Split SUSY:} Split SUSY was proposed in the aftermath of the emergence of
the landscape as a model which retained the desirable SUSY model feature of 
gauge coupling unification and a WIMP dark matter candidate whilst eschewing the requirement
of weak scale naturalness. In this approach, then, SUSY scalar masses other than the SM Higgs
can range from $\sim 10^{3}$ TeV up to possibly 
$10^{8}$ TeV\cite{ArkaniHamed:2004fb,Giudice:2004tc,ArkaniHamed:2004yi}. Meanwhile, gauginos and higgsinos are highly 
split from the scalars and can occupy masses in the $0.1-10$ TeV range. 
This type of split spectrum maintains gauge coupling unification and a mixed 
gaugino-higgsino type WIMP. 
The ultraheavy scalars suppress flavor- and CP-violating processes
and thus explain the lack of such effects in experiments. 
Recent work comparing split SUSY models
with the measured value of $m_h\simeq 125$ GeV favors a lower range of scalar masses
$\tm\sim 10-10^4$ TeV\cite{Giudice:2011cg}. Upon integrating out the supermassive scalar 
particles, then the low energy effective theory\cite{Delgado:2005ek} 
contains SM particles plus the gauginos 
and higgsinos. Thus, quadratic divergences should give rise to models as in Fig. \ref{fig:mhSM}
with a scale $\Lambda\sim \tm$: {\it i.e.} the required finetuning leads to a rare
occurance on the landscape for uniformly distributed values of $\mu_{SM}^{eff}$.

\item {\it High scale SUSY:} In High Scale (HS) SUSY, one again discards naturalness in the
hope that the landscape will solve the big hierarchy problem. 
However, for HS SUSY one allows {\it all} the SUSY partners to obtain large masses.
The energy scale for HS SUSY may be labeled as $\Lambda_{HS}$ and values considered in the
literature range from $10^2-10^{9}$ TeV\cite{HSSUSY}. Thus, the matching conditions
between the MSSM and the SM effective field theories are implemented around the
scale $\Lambda_{HS}$. The resultant Higgs mass dependence on $\mu_{SM}$ will be as in
Fig. \ref{fig:mhSM} so again we would expect HS SUSY to be a rare occurance on the
landscape as compared to RNS.

\item {\it PeV SUSY:} In PeV SUSY\cite{Wells:2004di}, SUSY breaking occurs via a charged
({\it i.e.} non-singlet) hidden sector $F$ term leading to scalar masses at the PeV scale
(1 PeV$=10^3$ TeV), whilst gaugino soft breaking terms are suppressed and hence dominated
by the anomaly-mediation form so that the wino turns out to be LSP. 
A 2-3 TeV wino is suggested to gain accord with the measured dark matter density
which in turn suggests $m_{3/2}\sim \tm\sim $ PeV. Some motivation for the PeV scale 
also comes from the neutrino sector by assuming neutrinos charged under an 
additional $U(1)^\prime$ so that Dirac neutrino masses arise from non-renormalizable operators.
The light Higgs mass is expected at $m_h\sim 125-145$ GeV. 
The PeV SUSY models would correspond to curves in Fig. \ref{fig:mhSM} with $\Lambda\sim 10^3$ TeV.
Thus, we would expect PeV SUSY to be relatively rare on the landscape.

\item {\it Spread SUSY:} In Spread SUSY\cite{Hall:2011jd}, three scales of sparticles
occur. The first possibility is that scalar masses occur at $\tm\sim 10^6$ TeV with
gauginos at $10^2$ TeV and higgsinos around $1$ TeV in order to gain accord with
the measured dark matter abundance. A second possibility considered has scalars
around $\tm\sim 10^3$ TeV with higgsinos an order of magnitude lower and
gauginos at the TeV scale (where winos with mass $\sim 3$ TeV would saturate the
measured dark matter abundance). Thus, Spread SUSY models would correspond to curves from
Fig. \ref{fig:mhSM} with $\Lambda\sim 10^3-10^6$ TeV. 
We would thus expect Spread SUSY to be rare on the string landscape.

\item {\it Minisplit:} Mini-split SUSY\cite{Arvanitaki:2012ps} 
emerged after the LHC Higgs discovery and attempted
to reconcile Split SUSY with a Higgs boson mass of $m_h\simeq 125$ GeV. 
To accommodate the Higgs mass, the mass scale of the heavy scalars was decreased to 
1. $\tm\sim 10^2$ TeV scale for a heavy $\mu\sim 10^2$ TeV model with 
TeV-scale gauginos (light AMSB) or 2. $\tm\sim 10^4$ TeV with $10^2$ TeV gauginos
but with small $\mu <$TeV (heavy AMSB). Models with $U(1)^\prime $ mediation were
also considered. The minisplit models would thus correspond to curves in Fig. \ref{fig:mhSM}
with $\Lambda\sim 10^2-10^4$ TeV. These models should be rare on the landscape, but
less rare than original Split SUSY models.
 

\end{itemize}

\section{Implications of stringy naturalness for collider and dark matter searches}
\label{sec:collDM}

In light of the implications of stringy naturalness, 
how then ought SUSY to be revealed at collider and dark matter search experiments?
Since the superpotential $\mu$ parameter must not be too far removed from 
our measured value of the weak scale, then higgsinos must also be light.
A compelling signature emerges from $\tchi_1^0\tchi_2^0$ production at 
LHC\cite{Baer:2011ec,SDLJMET} where
$\tchi_2^0\to\tchi_1^0\ell^+\ell^-$ in recoil against a hard radiated initial state jet.
The soft dilepton plus jet +MET signal should emerge slowly as more and more data accrues
at LHC\footnote{Excess events of this type seem to be emerging already in a recent Atlas
analysis with 139 fb$^{-1}$\cite{atlas:SDLJMET}.}.
Meanwhile, more conventional SUSY signatures such as those from gluino or top squark
pair production might be visible at HL-LHC\cite{Baer:2016wkz}, 
but also may require an upgrade to HE-LHC
since gluinos can range up to $m_{\tg}\alt 6$ TeV while stops are required at
$m_{\tst_1}\alt 3$ TeV\cite{Baer:2017pba,Baer:2018hpb}.
Same-sign diboson signatures from wino pair production are an additional 
possibility\cite{Baer:2013yha,Baer:2017gzf}.

The required light higgsinos within the natural mass range
$m(higgsinos)\sim 100-300$ GeV provide a lucrative target for an ILC-type
$e^+e^-$ collider with $\sqrt{s}>2m(higgsino)$. 
Such a machine would act as a higgsino factory where the various higgsino
masses could be precisely determined. 
The higgsino mass splittings are sensitive to the (heavier) gaugino masses; 
consequent fits to gaugino masses could then allow for 
tests of hypotheses regarding gaugino mass unification\cite{ilc}.

With regard to dark matter searches, we remark that naturalness in the QCD sector requires
the PQ solution to the strong CP problem and the concommitant axion $a$\cite{pqww}. 
To ensure $\bar{\theta}\alt 10^{-10}$, then the SUSY axion model must be safe from 
gravity corrections. 
This can occur when both PQ symmetry and R-parity conservation emerge
from underlying discrete $R$-symmetries, such as the recent example of 
$\mathbb{Z}_{24}^R$\cite{Z24R}.
In this case, one gains a solution to the SUSY $\mu$ problem\cite{mupaper} via a Kim-Nilles operator
leading to a SUSY DFSZ axion. While the SUSY DFSZ axion has suppressed couplings to
photons (and may not be observable with present technology\cite{axion}) the thermally 
underproduced higgsino-like WIMPs, 
which make up typically only about 10\% of dark matter, should ultimately be detectable
via multi-ton noble liquid detectors\cite{wimp}.

\section{Conclusions}
\label{sec:conclude}

In this paper we have explored Douglas' notion of stringy naturalness-- 
that the value of an observable ${\cal O}_2$ is more natural than ${\cal O}_1$ 
if the number of phenomenologically viable vacua giving rise to ${\cal O}_2$ 
is great than the number of vacua giving rise to ${\cal O}_1$-- 
and its relation to conventional naturalness, 
with regard to the big and little hierarchy problems and why the value of the weak scale 
$m_{weak}$ in {\it our} pocket universe is only $m_{weak}\sim 100$ GeV. 
We interpret {\it phenomenologically viable} to mean
a fertile patch of string theory vacua leading to the SM as the low energy effective theory
with a value for the weak scale not too far (a factor four) beyond our measured value, as
required by the nuclear physics calculations of Agrawal {\it et al.}\cite{Agrawal:1997gf}.

It is often claimed that the string theory landscape picture allows us to eschew the common
notion of naturalness in that certain parameters, such as the cosmological 
constant or the magnitude of the weak scale, are environmentally determined.
With regard to the SM, since there appears to be no theoretical preference for any value of
Higgs potential parameter $\mu_{SM}$, we would expect it to be roughly uniformly distributed
across the decades of possibilities. For the SM to be valid up to scales 
$\Lambda_{SM}\gg m_{weak}$, then only tiny permissible ranges of $\mu_{SM}$ could give rise 
to weak scale values leading to livable pocket universes (Fig. \ref{fig:mhSM}). 
In contrast, for effective theories like the MSSM, 
then since quadratic divergences all cancel, there appears to exist large swaths
of superpotential $\mu$ values leading to weak scales at $\sim 100$ GeV. 
Thus, we conclude that stringy naturalness would favor the MSSM over the SM 
as the appropriate low energy effective field theory, in agreement with, and not opposed to,
conventional notions of naturalness. 

We also explored implications of stringy naturalness for various SUSY models.
The CMSSM (mMSUGRA) model with $m_h\sim 125$ GeV-- where $\Delta_{EW}$ is found to be 
$\sim 10^2-10^4$-- should be rather infrequent on the landscape since only tiny ranges
of $\mu$ values lead to $m_{weak}\sim 100$ GeV. 
Likewise, the panoply of SUSY models with scalar masses in the $10^2-10^{11}$ TeV range 
(Split SUSY, HS SUSY, PeV SUSY, Spread SUSY and mini-split SUSY) all appear 
infrequent on the landscape due to the unlikelihood of vacua
which have unrelated parameters compensating for overly large contributions to the weak scale.
This is in spite of the rather general expectation that soft terms should be statistically
selected for large values, as expected in stringy models with multiple hidden sectors.

The statistical draw to large soft terms is just what is needed for SUSY with 
radiatively-driven naturalness, where one is {\it living dangerously} in that if the soft
terms are much larger, then one is placed into CCB or no EWSB vacua 
(which must be anthropically vetoed). This situation, that soft terms are expected as large 
as possible such that EW symmetry is barely broken and that all independent 
contributions to the weak scale are within a factor four of our measured value, is just what is
needed to radiatively drive the soft terms to natural values. 

In these radiative natural SUSY (RNS) models, we further compared the locus of stringy natural
regions of parameter space in the $m_0$ vs. $m_{1/2}$ plane to the conventionally natural regions. 
Here, there is a major difference: conventional naturalness favors soft terms
as close to the 100 GeV scale as possible while stringy naturalness favors soft terms
as large as possible such that the weak scale is not too far removed from our measured 
value of $\sim 100$ GeV. We can read off from Fig's \ref{fig:m0mhfn} that stringy naturalness
predicts a Higgs mass $m_h\sim 125$ GeV whilst sparticles remain (at present) beyond LHC reach.
Needless to say, this {\it postdiction} has been verified by Run 2 data from LHC13.

The stringy natural RNS SUSY model gives rise to specific tests at collider and dark matter
search experiment. We expect the soft dilepton plus jet plus MET signature to slowly emerge
at HL-LHC as more and more integrated luminosity accrues, while gluino, top squark
and wino pair production signals might require a HE-LHC for discovery. 
Stringy naturalness cries out for construction of an ILC $e^+e^-$ collider 
with $\sqrt{s}>2m(higgsino)$ which would act as a higgsino factory.
We still expect WIMPs at multi-ton noble liquid dark matter detection experiments but
SUSY DFSZ axions will likely be difficult to detect.

{\it Acknowledgements:} 
This work was supported in part by the US Department of Energy, Office
of High Energy Physics. 
The work of HB was performed at the Aspen Center for Physics, 
which is supported by National Science Foundation grant PHY-1607611.


%

\begin{thebibliography}{99}

\bibitem{ghp} L.~Susskind,
``Dynamics of Spontaneous Symmetry Breaking in the Weinberg-Salam Theory,''
  Phys.\ Rev.\ D {\bf 20}, 2619 (1979).
%
\bibitem{witten} E.~Witten,
``Dynamical Breaking of Supersymmetry,''
  Nucl.\ Phys.\ B {\bf 188}, 513 (1981);
R.~K.~Kaul,
``Gauge Hierarchy in a Supersymmetric Model,''
  Phys.\ Lett.\  {\bf 109B}, 19 (1982).
%
\bibitem{WSS} H.~Baer and X.~Tata,
  ``Weak scale supersymmetry: From superfields to scattering events,''
  Cambridge, UK: Univ. Pr. (2006) 537 p.;
for a recent review, see {\it e.g.} 
D.~J.~H.~Chung, L.~L.~Everett, G.~L.~Kane, S.~F.~King, J.~D.~Lykken and L.~T.~Wang,
``The Soft supersymmetry breaking Lagrangian: Theory and applications,''
  Phys.\ Rept.\  {\bf 407}, 1 (2005).
%
\bibitem{Xe1ton} E.~Aprile {\it et al.} [XENON Collaboration],
``Dark Matter Search Results from a One Ton-Year Exposure of XENON1T,''
  Phys.\ Rev.\ Lett.\  {\bf 121}, no. 11, 111302 (2018);
X.~Cui {\it et al.} [PandaX-II Collaboration],
``Dark Matter Results From 54-Ton-Day Exposure of PandaX-II Experiment,''
  Phys.\ Rev.\ Lett.\  {\bf 119}, no. 18, 181302 (2017);
D.~S.~Akerib {\it et al.} [LUX Collaboration],
``Results from a search for dark matter in the complete LUX exposure,''
  Phys.\ Rev.\ Lett.\  {\bf 118}, no. 2, 021303 (2017).
%
\bibitem{lhc_gl} M.~Aaboud {\it et al.} [ATLAS Collaboration],
 ``Search for squarks and gluinos in final states with jets and missing transverse momentum using 36  fb$^{-1}$ of $\sqrt{s}=13$  TeV pp collision data with the ATLAS detector,''
  Phys.\ Rev.\ D {\bf 97}, no. 11, 112001 (2018).
%
\bibitem{lhc_t1} A.~M.~Sirunyan {\it et al.} [CMS Collaboration],
``Search for top squark pair production in pp collisions at $ \sqrt{s}=13 $ TeV using single lepton events,''
  JHEP {\bf 1710}, 019 (2017);

The ATLAS collaboration [ATLAS Collaboration],
``Search for direct top squark pair production in the 3-body decay mode with a final state containing one lepton, jets, and missing transverse momentum in $\sqrt{s}=13$TeV $pp$ collision data with the ATLAS detector,''
  ATLAS-CONF-2019-017.
%
\bibitem{EENZ} J.~R.~Ellis, K.~Enqvist, D.~V.~Nanopoulos and F.~Zwirner,
``Observables in Low-Energy Superstring Models,''
  Mod.\ Phys.\ Lett.\ A {\bf 1}, 57 (1986).
%
\bibitem{BG} R.~Barbieri and G.~F.~Giudice,
``Upper Bounds on Supersymmetric Particle Masses,''
Nucl.\ Phys.\ B {\bf 306}, 63 (1988).
%
\bibitem{DG} S.~Dimopoulos and G.~F.~Giudice,
``Naturalness constraints in supersymmetric theories with nonuniversal soft terms,''
  Phys.\ Lett.\ B {\bf 357}, 573 (1995).
%
\bibitem{AC} G.~W.~Anderson and D.~J.~Castano,
``Challenging weak scale supersymmetry at colliders,''
Phys.\ Rev.\ D {\bf 53}, 2403 (1996).
%
\bibitem{Kim:2013uxa} D.~Kim, P.~Athron, C.~Balázs, B.~Farmer and E.~Hutchison,
``Bayesian naturalness of the CMSSM and CNMSSM,''
  Phys.\ Rev.\ D {\bf 90}, no. 5, 055008 (2014);
P.~Athron, C.~Balazs, B.~Farmer, A.~Fowlie, D.~Harries and D.~Kim,
``Bayesian analysis and naturalness of (Next-to-)Minimal Supersymmetric Models,''
  JHEP {\bf 1710}, 160 (2017).
%
\bibitem{BS} R.~Barbieri and A.~Strumia,
``About the fine tuning price of LEP,''
  Phys.\ Lett.\ B {\bf 433}, 63 (1998).
%
\bibitem{land3} H.~Baer, V.~Barger, S.~Salam, H.~Serce and K.~Sinha,
``LHC SUSY and WIMP dark matter searches confront the string theory landscape,''
  JHEP {\bf 1904}, 043 (2019).
%
\bibitem{dew} H.~Baer, V.~Barger and D.~Mickelson,
``How conventional measures overestimate electroweak fine-tuning in supersymmetric theory,''
Phys.\ Rev.\ D {\bf 88}, 095013 (2013).
%
\bibitem{mt} A.~Mustafayev and X.~Tata,
``Supersymmetry, Naturalness, and Light Higgsinos,''
  Indian J.\ Phys.\  {\bf 88}, 991 (2014).
%
\bibitem{seige} H.~Baer, V.~Barger, D.~Mickelson and M.~Padeffke-Kirkland,
``SUSY models under siege: LHC constraints and electroweak fine-tuning,''
Phys.\ Rev.\ D {\bf 89}, 115019 (2014).
%
\bibitem{arno} H.~Baer, V.~Barger and M.~Savoy,
``Supergravity gauge theories strike back: There is no crisis for SUSY but a new collider may be required for discovery,''
  Phys.\ Scripta {\bf 90}, 068003 (2015).
%
\bibitem{prw} M.~Papucci, J.~T.~Ruderman and A.~Weiler,
``Natural SUSY Endures,''
  JHEP {\bf 1209}, 035 (2012).
%
\bibitem{bkls} C.~Brust, A.~Katz, S.~Lawrence and R.~Sundrum,
``SUSY, the Third Generation and the LHC,''
  JHEP {\bf 1203}, 103 (2012).
%
\bibitem{mhiggs} M.~Carena and H.~E.~Haber,
``Higgs boson theory and phenomenology,''
  Prog.\ Part.\ Nucl.\ Phys.\  {\bf 50}, 63 (2003);
P.~Draper and H.~Rzehak,
``A Review of Higgs Mass Calculations in Supersymmetric Models,''
  Phys.\ Rept.\  {\bf 619}, 1 (2016).
%
\bibitem{h125} H.~Baer, V.~Barger and A.~Mustafayev,
``Implications of a 125 GeV Higgs scalar for LHC SUSY and neutralino dark matter searches,''
  Phys.\ Rev.\ D {\bf 85}, 075010 (2012).
%
\bibitem{ltr} H.~Baer, V.~Barger, P.~Huang, A.~Mustafayev and X.~Tata,
``Radiative natural SUSY with a 125~GeV Higgs boson,''
Phys. Rev. Lett. {\bf 109}, 161802 (2012).
%
\bibitem{rns} H.~Baer, V.~Barger, P.~Huang, D.~Mickelson, A.~Mustafayev and X.~Tata,
``Radiative natural supersymmetry: Reconciling electroweak fine-tuning and the Higgs boson mass,''
Phys.\ Rev.\ D {\bf 87}, 115028 (2013).
%
\bibitem{upper} H.~Baer, V.~Barger and M.~Savoy,
``Upper bounds on sparticle masses from naturalness or how to disprove weak scale supersymmetry,''
  Phys.\ Rev.\ D {\bf 93}, no. 3, 035016 (2016).
%
\bibitem{jamie} H.~Baer, V.~Barger, J.~S.~Gainer, P.~Huang, M.~Savoy, H.~Serce and X.~Tata,
``What hadron collider is required to discover or falsify natural supersymmetry?,''
  Phys.\ Lett.\ B {\bf 774}, 451 (2017).
%
\bibitem{maren} H.~Baer, V.~Barger, M.~Padeffke-Kirkland and X.~Tata,
``Naturalness implies intra-generational degeneracy for decoupled squarks and sleptons,''
  Phys.\ Rev.\ D {\bf 89},  no.3, 037701(2014).
%
\bibitem{Weinberg:1987dv}
  S.~Weinberg,
``Anthropic Bound on the Cosmological Constant,''
  Phys.\ Rev.\ Lett.\  {\bf 59}, 2607 (1987);
S.~Weinberg,
``The Cosmological Constant Problem,''
  Rev.\ Mod.\ Phys.\  {\bf 61}, 1 (1989).
%
\bibitem{BP}   R.~Bousso and J.~Polchinski,
``Quantization of four form fluxes and dynamical neutralization of the cosmological constant,''
  JHEP {\bf 0006},006 (2000).
%
\bibitem{Susskind:2003kw}
  L.~Susskind,
``The Anthropic landscape of string theory,''
  In *Carr, Bernard (ed.): Universe or multiverse?* 247-266
  [hep-th/0302219].
%
\bibitem{Douglas:2004qg} M.~R.~Douglas,
``Statistical analysis of the supersymmetry breaking scale,''
  hep-th/0405279.
%
\bibitem{Ashok:2003gk} S.~Ashok and M.~R.~Douglas,
``Counting flux vacua,''
  JHEP {\bf 0401}, 060 (2004).
%
\bibitem{ArkaniHamed:2004fb} N.~Arkani-Hamed and S.~Dimopoulos,
``Supersymmetric unification without low energy supersymmetry and signatures for fine-tuning at the LHC,''
  JHEP {\bf 0506}, 073 (2005).
%
\bibitem{HSSUSY} V.~Barger, C.~W.~Chiang, J.~Jiang and T.~Li,
``Axion models with high-scale supersymmetry breaking,''
  Nucl.\ Phys.\ B {\bf 705}, 71 (2005);
V.~Barger, J.~Jiang, P.~Langacker and T.~Li,
``Non-canonical gauge coupling unification in high-scale supersymmetry breaking,''
  Nucl.\ Phys.\ B {\bf 726}, 149 (2005);
C.~Liu,
``A supersymmetry model of leptons,''
  Phys.\ Lett.\ B {\bf 609}, 111 (2005);
C.~Liu,
``Supersymmetry for fermion masses,''
  Commun.\ Theor.\ Phys.\  {\bf 47}, 1088 (2007);
C.~Liu and Z.~h.~Zhao,
``$\theta_{13}$ and the Higgs mass from high scale supersymmetry,''
  Commun.\ Theor.\ Phys.\  {\bf 59}, 467 (2013);
G.~Elor, H.~S.~Goh, L.~J.~Hall, P.~Kumar and Y.~Nomura,
``Environmentally Selected WIMP Dark Matter with High-Scale Supersymmetry Breaking,''
  Phys.\ Rev.\ D {\bf 81}, 095003 (2010);
R.~Sato, S.~Shirai and K.~Tobioka,
``Gluino Decay as a Probe of High Scale Supersymmetry Breaking,''
  JHEP {\bf 1211}, 041 (2012);
J.~Hisano, K.~Ishiwata and N.~Nagata,
``Direct Search of Dark Matter in High-Scale Supersymmetry,''
  Phys.\ Rev.\ D {\bf 87},  035020 (2013);
J.~Hisano, T.~Kuwahara and N.~Nagata,
``Grand Unification in High-scale Supersymmetry,''
  Phys.\ Lett.\ B {\bf 723}, 324 (2013);
I.~Masina and M.~Quiros,
``On the Veltman Condition, the Hierarchy Problem and High-Scale Supersymmetry,''
  Phys.\ Rev.\ D {\bf 88},  093003 (2013);
N.~Nagata and S.~Shirai,
``Higgsino Dark Matter in High-Scale Supersymmetry,''
  JHEP {\bf 1501}, 029 (2015);
S.~A.~R.~Ellis and J.~D.~Wells,
``High-scale supersymmetry, the Higgs boson mass, and gauge unification,''
  Phys.\ Rev.\ D {\bf 96}, no.5,  055024 no.11,  115032; (2017);
E.~Dudas, T.~Gherghetta, Y.~Mambrini and K.~A.~Olive,
``Inflation and High-Scale Supersymmetry with an EeV Gravitino,''
  Phys.\ Rev.\ D {\bf 96}, no. 11, 115032 (2017);
S.~A.~R.~Ellis, T.~Gherghetta, K.~Kaneta and K.~A.~Olive,
``New Weak-Scale Physics from SO(10) with High-Scale Supersymmetry,''
  Phys.\ Rev.\ D {\bf 98},  no.5,  055009 (2018);
K.~Y.~Choi and H.~M.~Lee,
``Axino abundances in high-scale supersymmetry,''
  Phys.\ Dark Univ.\  {\bf 22}, 202 (2018);
E.~Dudas, T.~Gherghetta, K.~Kaneta, Y.~Mambrini and K.~A.~Olive,
``Limits on $R$-parity Violation in High Scale Supersymmetry,''
  arXiv:1905.09243 [hep-ph].
%
\bibitem{Wells:2004di}
  J.~D.~Wells,
``PeV-scale supersymmetry,''
  Phys.\ Rev.\ D {\bf 71}, 015013 (2005).
%
\bibitem{Hall:2011jd}
  L.~J.~Hall and Y.~Nomura,
``Spread Supersymmetry,''
  JHEP {\bf 1201}, 082 (2012);
L.~J.~Hall, Y.~Nomura and S.~Shirai,
``Spread Supersymmetry with Wino LSP: Gluino and Dark Matter Signals,''
  JHEP {\bf 1301}, 036 (2013).
%
\bibitem{Arvanitaki:2012ps}
  A.~Arvanitaki, N.~Craig, S.~Dimopoulos and G.~Villadoro,
``Mini-Split,''
  JHEP {\bf 1302}, 126 (2013).
%
\bibitem{Douglas:2004zg}
  M.~R.~Douglas,
``Basic results in vacuum statistics,''
  Comptes Rendus Physique {\bf 5}, 965 (2004).
%
\bibitem{Agrawal:1997gf}  V.~Agrawal, S.~M.~Barr, J.~F.~Donoghue and D.~Seckel,
``The Anthropic principle and the mass scale of the standard model,''
  Phys.\ Rev.\ D {\bf 57}, 5480 (1998);
V.~Agrawal, S.~M.~Barr, J.~F.~Donoghue and D.~Seckel,
``Anthropic considerations in multiple domain theories and the scale of electroweak symmetry breaking,''
  Phys.\ Rev.\ Lett.\  {\bf 80}, 1822 (1998).
%
\bibitem{Baer:2016lpj}
  H.~Baer, V.~Barger, M.~Savoy and H.~Serce,
``The Higgs mass and natural supersymmetric spectrum from the landscape,''
  Phys.\ Lett.\ B {\bf 758}, 113 (2016).
%
\bibitem{land} H.~Baer, V.~Barger, H.~Serce and K.~Sinha,
 ``Higgs and superparticle mass predictions from the landscape,''
  JHEP {\bf 1803}, 002 (2018).
%
\bibitem{Baer:2019uom}
  H.~Baer, V.~Barger, D.~Sengupta, H.~Serce, K.~Sinha and R.~W.~Deal,
``Is the magnitude of the Peccei-Quinn scale set by the landscape?,''
  arXiv:1905.00443 [hep-ph].
%
\bibitem{Dienes:2007ms}
  K.~R.~Dienes, M.~Lennek, D.~Senechal and V.~Wasnik,
``Supersymmetry versus Gauge Symmetry on the Heterotic Landscape,''
  Phys.\ Rev.\ D {\bf 75}, 126005 (2007);
K.~R.~Dienes, M.~Lennek, D.~Senechal and V.~Wasnik,
``Is SUSY Natural?,''
  New J.\ Phys.\  {\bf 10}, 085003 (2008).
%
\bibitem{suss} L.~Susskind,
``Supersymmetry breaking in the anthropic landscape,''
  In *Shifman, M. (ed.) et al.: From fields to strings, vol. 3* 1745-1749
  [hep-th/0405189].
%
\bibitem{denefdouglas}
  F.~Denef and M.~R.~Douglas,
``Distributions of flux vacua,''
  JHEP {\bf 0405}, 072 (2004).
%
\bibitem{mupaper} K.~J.~Bae, H.~Baer, V.~Barger and D.~Sengupta,
``Revisiting the SUSY $\mu$ problem and its solutions in the LHC era,''
  Phys.\ Rev.\ D {\bf 99}, no. 11, 115027 (2019).
%
\bibitem{Z24R} H.~Baer, V.~Barger and D.~Sengupta,
 ``Gravity safe, electroweak natural axionic solution to strong $CP$ and SUSY $\mu$ problems,''
  Phys.\ Lett.\ B {\bf 790}, 58 (2019).
%
\bibitem{KN} J.~E.~Kim and H.~P.~Nilles,
``The mu Problem and the Strong CP Problem,''
  Phys.\ Lett.\  {\bf 138B}, 150 (1984).
%
\bibitem{lrrrssv2} H.~M.~Lee, S.~Raby, M.~Ratz, G.~G.~Ross, R.~Schieren, K.~Schmidt-Hoberg and P.~K.~S.~Vaudrevange,
``Discrete R symmetries for the MSSM and its singlet extensions,''
  Nucl.\ Phys.\ B {\bf 850}, 1 (2011).
%
\bibitem{cmssm} For a review, see {\it e.g.}
R.~Arnowitt and P.~Nath,
``Developments in Supergravity Unified Models,''                                                   
  In *Kane, G.L. (ed.): Perspectives on supersymmetry II* 222-243
  [arXiv:0912.2273 [hep-ph]]  and references therein;
V.~D.~Barger, M.~S.~Berger and P.~Ohmann,
``Supersymmetric grand unified theories: Two loop evolution of gauge and Yukawa couplings,''
  Phys.\ Rev.\ D {\bf 47}, 1093 (1993) and 
  Phys.\ Rev.\ D {\bf 49}, 4908 (1994);
G.~L.~Kane, C.~F.~Kolda, L.~Roszkowski and J.~D.~Wells,
``Study of constrained minimal supersymmetry,''
Phys.\ Rev.\ D {\bf 49}, 6173 (1994).
%
 \bibitem{isajet} F.~E.~Paige, S.~D.~Protopopescu, H.~Baer and X.~Tata,
``ISAJET 7.69: A Monte Carlo event generator for pp, anti-p p, and e+e- reactions,''
  hep-ph/0312045.
%
\bibitem{ccn} K.~L.~Chan, U.~Chattopadhyay and P.~Nath,
``Naturalness, weak scale supersymmetry and the prospect for the observation 
of supersymmetry at the Tevatron and at the CERN LHC,''                           Phys.\ Rev.\ D {\bf 58}, 096004 (1998).
%
\bibitem{fmm} J.~L.~Feng, K.~T.~Matchev and T.~Moroi,
``Multi - TeV scalars are natural in minimal supergravity,''
  Phys.\ Rev.\ Lett.\  {\bf 84}, 2322 (2000);
J.~L.~Feng, K.~T.~Matchev and T.~Moroi,
``Focus points and naturalness in supersymmetry,''
  Phys.\ Rev.\ D {\bf 61}, 075005 (2000).
%
\bibitem{nuhm2}  D.~Matalliotakis and H.~P.~Nilles,
``Implications of nonuniversality of soft terms in supersymmetric grand unified theories,''
  Nucl.\ Phys.\ B {\bf 435}, 115 (1995);
M.~Olechowski and S.~Pokorski,
``Electroweak symmetry breaking with nonuniversal scalar soft terms and large tan beta solutions,''
  Phys.\ Lett.\ B {\bf 344}, 201 (1995);
P.~Nath and R.~L.~Arnowitt,
``Nonuniversal soft SUSY breaking and dark matter,''
  Phys.\ Rev.\ D {\bf 56}, 2820 (1997);
J.~R.~Ellis, K.~A.~Olive and Y.~Santoso,
``The MSSM parameter space with nonuniversal Higgs masses,''
  Phys.\ Lett.\ B {\bf 539}, 107 (2002);
J.~R.~Ellis, T.~Falk, K.~A.~Olive and Y.~Santoso,
``Exploration of the MSSM with nonuniversal Higgs masses,''
  Nucl.\ Phys.\ B {\bf 652}, 259 (2003);
H.~Baer, A.~Mustafayev, S.~Profumo, A.~Belyaev and X.~Tata,
``Direct, indirect and collider detection of neutralino dark matter in SUSY models with non-universal Higgs masses,''
  JHEP {\bf 0507}, 065 (2005).
%
\bibitem{miniland} W.~Buchmuller, K.~Hamaguchi, O.~Lebedev and M.~Ratz,
``Supersymmetric standard model from the heterotic string,''
  Phys.\ Rev.\ Lett.\  {\bf 96}, 121602 (2006);
W.~Buchmuller, K.~Hamaguchi, O.~Lebedev and M.~Ratz,
``Supersymmetric Standard Model from the Heterotic String (II),''
  Nucl.\ Phys.\ B {\bf 785}, 149 (2007);
O.~Lebedev, H.~P.~Nilles, S.~Raby, S.~Ramos-Sanchez, M.~Ratz, P.~K.~S.~Vaudrevange and A.~Wingerter,
``A Mini-landscape of exact MSSM spectra in heterotic orbifolds,''
  Phys.\ Lett.\ B {\bf 645}, 88 (2007);
O.~Lebedev, H.~P.~Nilles, S.~Raby, S.~Ramos-Sanchez, M.~Ratz, P.~K.~S.~Vaudrevange and A.~Wingerter,
``The Heterotic Road to the MSSM with R parity,''
  Phys.\ Rev.\ D {\bf 77}, 046013 (2008);
O.~Lebedev, H.~P.~Nilles, S.~Ramos-Sanchez, M.~Ratz and P.~K.~S.~Vaudrevange,
``Heterotic mini-landscape. (II). Completing the search for MSSM vacua in a Z(6) orbifold,''
  Phys.\ Lett.\ B {\bf 668}, 331 (2008); 
for a review, see H.~P.~Nilles and P.~K.~S.~Vaudrevange,
``Geography of Fields in Extra Dimensions: String Theory Lessons for Particle Physics,''
  Mod.\ Phys.\ Lett.\ A {\bf 30}, no.10, 1530008 (2015).
%
\bibitem{ArkaniHamed:2005yv}
  N.~Arkani-Hamed, S.~Dimopoulos and S.~Kachru,
``Predictive landscapes and new physics at a TeV,''
  hep-th/0501082.
%
\bibitem{Giudice:2006sn}
  G.~F.~Giudice and R.~Rattazzi,
``Living Dangerously with Low-Energy Supersymmetry,''
  Nucl.\ Phys.\ B {\bf 757}, 19 (2006).
%
\bibitem{Giudice:2004tc}
  G.~F.~Giudice and A.~Romanino,
``Split supersymmetry,''
  Nucl.\ Phys.\ B {\bf 699}, 65 (2004)
   Erratum: [Nucl.\ Phys.\ B {\bf 706}, 487 (2005).
%
\bibitem{ArkaniHamed:2004yi}
  N.~Arkani-Hamed, S.~Dimopoulos, G.~F.~Giudice and A.~Romanino,
``Aspects of split supersymmetry,''
  Nucl.\ Phys.\ B {\bf 709}, 3 (2005).
%
\bibitem{Giudice:2011cg}
  G.~F.~Giudice and A.~Strumia,
``Probing High-Scale and Split Supersymmetry with Higgs Mass Measurements,''
  Nucl.\ Phys.\ B {\bf 858}, 63 (2012).
%
\bibitem{Delgado:2005ek}
  A.~Delgado and G.~F.~Giudice,
``On the tuning condition of split supersymmetry,''
  Phys.\ Lett.\ B {\bf 627}, 155 (2005).
%
\bibitem{Baer:2011ec}
  H.~Baer, V.~Barger and P.~Huang,
``Hidden SUSY at the LHC: the light higgsino-world scenario and the role of a lepton collider,''
  JHEP {\bf 1111}, 031 (2011).
%
\bibitem{SDLJMET} Z.~Han, G.~D.~Kribs, A.~Martin and A.~Menon,
``Hunting quasidegenerate Higgsinos,''
Phys.\ Rev.\ D {\bf 89}, no.7,  075007 (2014);
H.~Baer, A.~Mustafayev and X.~Tata,
``Monojet plus soft dilepton signal from light higgsino pair production at LHC14,''                
  Phys.\ Rev.\ D {\bf 90}, no.11,  115007 (2014);
C.~Han, D.~Kim, S.~Munir and M.~Park,
``Accessing the core of naturalness, nearly degenerate higgsinos, at the LHC,''                    
  JHEP {\bf 1504}, 132 (2015);
H.~Baer, V.~Barger, M.~Savoy and X.~Tata,
``Multichannel assault on natural supersymmetry at the high luminosity LHC,''
  Phys.\ Rev.\ D {\bf 94}, no.3,  035025 (2016).
%
\bibitem{atlas:SDLJMET} The ATLAS collaboration [ATLAS Collaboration],
``Searches for electroweak production of supersymmetric particles with compressed mass spectra in $\sqrt{s} = 13$ TeV $pp$ collisions with the ATLAS detector,''
  ATLAS-CONF-2019-014.
%
\bibitem{Baer:2016wkz}
  H.~Baer, V.~Barger, J.~S.~Gainer, P.~Huang, M.~Savoy, D.~Sengupta and X.~Tata,
``Gluino reach and mass extraction at the LHC in radiatively-driven natural SUSY,''
  Eur.\ Phys.\ J.\ C {\bf 77}, no.7,  499 (2017).
%
\bibitem{Baer:2017pba}
  H.~Baer, V.~Barger, J.~S.~Gainer, H.~Serce and X.~Tata,
``Reach of the high-energy LHC for gluinos and top squarks in SUSY models with light Higgsinos,''
  Phys.\ Rev.\ D {\bf 96}, no.11,  115008 (2017).
%
\bibitem{Baer:2018hpb}
  H.~Baer, V.~Barger, J.~S.~Gainer, D.~Sengupta, H.~Serce and X.~Tata,
``LHC luminosity and energy upgrades confront natural supersymmetry models,''
  Phys.\ Rev.\ D {\bf 98}, no.7,  075010 (2018).
%
\bibitem{Baer:2013yha}
  H.~Baer, V.~Barger, P.~Huang, D.~Mickelson, A.~Mustafayev, W.~Sreethawong and X.~Tata,
``Same sign diboson signature from supersymmetry models with light higgsinos at the LHC,''
  Phys.\ Rev.\ Lett.\  {\bf 110}, no.15,  151801 (2013).
%
\bibitem{Baer:2017gzf}
  H.~Baer, V.~Barger, J.~S.~Gainer, M.~Savoy, D.~Sengupta and X.~Tata,
``Aspects of the same-sign diboson signature from wino pair production with light higgsinos at the high luminosity LHC,''
  Phys.\ Rev.\ D {\bf 97}, no.3,  035012 (2018).
%
\bibitem{ilc} H.~Baer, V.~Barger, D.~Mickelson, A.~Mustafayev and X.~Tata,
``Physics at a Higgsino Factory,''
  JHEP {\bf 1406}, 172 (2014);
S.~L.~Lehtinen, H.~Baer, M.~Berggren, K.~Fujii, J.~List, T.~Tanabe and J.~Yan,
``Naturalness and light Higgsinos: why ILC is the right machine for SUSY discovery,''
  arXiv:1710.02406 [hep-ph];
H.~Baer, M.~Berggren, K.~Fujii, J.~List, S.~L.~Lehtinen, T.~Tanabe and J.~Yan,
to appear.
%
\bibitem{pqww} R.~D.~Peccei and H.~R.~Quinn,
``CP Conservation in the Presence of Instantons,''
  Phys.\ Rev.\ Lett.\  {\bf 38}, 1440 (1977);
R.~D.~Peccei and H.~R.~Quinn,
``Constraints Imposed by CP Conservation in the Presence of Instantons,''
  Phys.\ Rev.\ D {\bf 16}, 1791 (1977);
S.~Weinberg,
``A New Light Boson?,''
  Phys.\ Rev.\ Lett.\  {\bf 40}, 223 (1978);
F.~Wilczek,
``Problem of Strong  $P$  and  $T$  Invariance in the Presence of Instantons,''
  Phys.\ Rev.\ Lett.\  {\bf 40}, 279 (1978). 
%
\bibitem{axion} K.~J.~Bae, H.~Baer and H.~Serce,
``Prospects for axion detection in natural SUSY with mixed axion-higgsino dark matter: back to invisible?,''
  JCAP {\bf 1706}, no.06,  024 (2017).
%
\bibitem{wimp} H.~Baer, V.~Barger and D.~Mickelson,
``Direct and indirect detection of higgsino-like WIMPs: concluding the story of electroweak naturalness,''
  Phys.\ Lett.\ B {\bf 726}, 330 (2013);
H.~Baer, V.~Barger and H.~Serce,
``SUSY under siege from direct and indirect WIMP detection experiments,''
  Phys.\ Rev.\ D {\bf 94}, no.11,  115019 (2016).
%
\end{thebibliography}
\end{document}